\documentclass[twocolumn]{aastex631}
\usepackage{natbib}
\usepackage{hyperref}
\usepackage{breakurl}
\usepackage{mathtools}
\usepackage{graphicx}

\shorttitle{Variation of coronal size in a jetted TDE}
\shortauthors{Chatterjee et al.}
\graphicspath{{./}{figures/}}

\newcommand{\swift}{{\it Swift} }
\newcommand{\xmm}{{\it XMM-Newton} }
\newcommand{\powerlaw}{{\tt powerlaw} }

\newcommand{\ecs}{erg cm$^{-2}$ s$^{-1}$ }

\newcommand{\source}{{\it Swift} J1644+57}
\newcommand{\M}{$M_{\odot}$ }
\newcommand{\phc}{ph cm$^{-2}$ s$^{-1}$ }

\usepackage{color}

\begin{document}

\title{Temporal Variation of the Coronal Radius Parameter in a Jetted Tidal Disruption Event: Swift J1644+57}

\correspondingauthor{Arka Chatterjee}
\email{arka.chatterjee@ddn.upes.ac.in}

\author[0000-0003-3932-6705]{Arka Chatterjee}
\affiliation{Department of Physics, School of Advanced Engineering, UPES, Dehradun, 248007, India\\}
\affiliation{Department of Physics \& Astronomy, Faculty of Science, University of Manitoba, Winnipeg, Manitoba, R3T 2N2, Canada\\}

\author[0000-0003-4799-1895]{Kimitake Hayasaki}
\affiliation{Department of Astronomy and Space Science, Chungbuk National University, Cheongju, 361-763, Republic of Korea\\}
\affiliation{Department of Physical Sciences, Aoyama Gakuin University, Sagamihara 252-5258, Japan\\}

\author[0000-0003-3840-0571]{Prantik Nandi}
\affiliation{Physical Research Laboratory, Navrangpura, Ahmedabad, 380009, India\\}
\affiliation{Indian Centre for Space Physics, Barakhola, Netai Nagar, Kolkata, 700099 India\\}

\author[0000-0003-0071-8947]{Neeraj Kumari}
\affiliation{Indian Institute of Astrophysics, Koramangala, Bengaluru 560 034, India\\}
\affiliation{INAF-IASF Palermo, Via Ugo La Malfa 153, I-90146 Palermo, Italy\\}

\author[0000-0001-5294-7667]{Skye R. Heiland}
\affiliation{Department of Physics \& Astronomy, University of British Columbia, Vancouver, BC, V6T 1Z4, Canada\\}

\author[0000-0001-7500-5752]{Arghajit Jana}
\affiliation{Department of Physics, SRM University-AP, Amaravati, Andhra Pradesh 522240, India \\}

\author[0000-0003-2865-4666]{Sachindra Naik}
\affiliation{Physical Research Laboratory, Navrangpura, Ahmedabad, 380009, India\\}

\author[0000-0001-6189-7665]{Samar Safi-Harb}
\affiliation{Department of Physics \& Astronomy, Faculty of Science, University of Manitoba, Winnipeg, Manitoba, R3T 2N2, Canada\\}

\begin{abstract}
Tidal Disruption Events are exotic astrophysical phenomena where matter from a star or 
the interstellar medium is captured by a supermassive black hole. The process liberates 
enormous energy, within a few months to a year timescale, enough to detect dormant black 
holes in near as well as the farthest galaxies. We revisit the long-term spectral variabilities associated with the jetted Tidal Disruption Event \source~by exploring the archival X-ray data obtained with Swift/XRT and XMM-Newton observatories. Our analysis reveals that the spectral indices decrease non-monotonically as \source~evolves with time. We also find that the soft (0.3-1.5 keV) and hard (1.5-10 keV) X-ray photon counts are highly correlated with a maximum correlation coefficient of 0.95 and peak at {\it zero} lag. Moreover, the soft and hard band variabilities obtained from XMM-Newton observations are highly correlated with a Pearson cross-correlation coefficient of 0.96. This indicates that the soft and hard X-ray photons are emitted from the same site, which is most likely a Compton cloud, i.e., the corona. Assuming the hard X-ray photons originate from the corona, we find that the coronal parameter undergoes rapid expansion during the early phases when accompanied by a relativistic jet launching and subsequently evolves toward a state of saturation with minor fluctuations in the latter stages. The temporal variation of the coronal radius parameter ($R_{cor}$) is consistent with a simple theoretical conjecture. We also discuss the application of our analytical outcomes to other jetted and non-jetted tidal disruption events.
\end{abstract}


%
\section{Introduction} 
\label{sec:intro}
%
Most galaxies are thought to harbor supermassive black holes (SMBHs) within the mass range of $10^6\, M_\odot\lesssim{M_{\rm bh}}\lesssim10^{10}\, M_\odot$ \citep{KR1995, KH2013}. The mass of the SMBH has been estimated using the reverberation mapping technique \citep{Peterson2004}, measuring the proper motion of stars orbiting around the SMBH, and the $M-\sigma_*$ relation, where the velocity dispersion of stars ($\sigma_*$) around the galactic bulge is correlated with the mass of the SMBH ($M$). 

Tidal disruption events (TDEs) can provide a unique way to probe the dormant SMBHs in inactive galaxies. A star is tidally disrupted when it reaches within the tidal disruption radius $R_{\rm t}=(M_{\rm bh}/m_*)r_*$, where $M_{\rm bh}$ is the mass of the SMBH, $m_*$ is the stellar mass, and $r_*$ is the stellar radius. 

The tidal disruption radius is expressed by

\begin{equation}
\frac{R_{\rm t}}{r_{\rm g}}
\approx
50\,
\left(\frac{M_{\rm bh}}{10^6\,M_{\odot}}\right)^{-2/3}
\left(\frac{m_{*}}{\,M_{\odot}}\right)^{-1/3}
\left(\frac{r_{*}}{\,R_{\odot}}\right)
\label{eq:1}
\end{equation}
in units of the gravitational radius,
$r_{\rm g}= GM_{\rm bh}/c^2$, where is $G$ is the gravitational constant, 
$M_{\rm bh}$ is the mass of the black hole, and $c$ is the speed of light.

After tidal disruption of the star, the stellar debris falls toward the SMBH and releases its gravitational energy as radiation, generating a luminosity large enough to be comparable to the Eddington limit. According to the standard TDE theory (\citealt{Rees1988}; see also \citealt{Rossi+2021} for a review), the resultant luminosity decays with time, following the power law of time $t^{-5/3}$, for a span of a few 100 days over the wide range of optical/UV to X-ray wavebands. It is a key observational feature that the light curve decays with $t^{-5/3}$. Though a significant number of TDEs can deviate from the standard $t^{-5/3}$ decay rate due to stellar internal structure \citep{2011Lodato}, orbital eccentricity \citep{Hayasaki2013, Hayasaki2016, Hayasaki2018, PGKH2020,2023ApJ...959...19Z}, and/or the penetration factor which is the ratio of the tidal disruption to pericenter radii \citep{James2013}.

The first observational evidence of TDEs dates back to the 1990s \citep{Bade1996} by the All-Sky Survey of the {\it ROSAT} satellite. Since then, through multi-wavelength campaigns, $\sim100$ \citep{Gezari2021} have been discovered in a wide range of wavebands from radio to optical/UV to X-ray bands. While most TDEs have been observed in optical/UV wavelengths, $\sim10$ TDEs have been detected simultaneously in the UV/optical and soft X-ray wavelengths. In addition to these thermal TDEs without a relativistic jet (so-called non-jetted TDEs), there have been so far four remarkable TDEs that accompany a relativistic jet (jetted TDEs), such as {\it Swift}~J164449.3+573451 (hereafter, {\it Swift}~J1644+57) \citep{Levan2011, Bloom2011a, Bloom2011b, Burrows2011}, {\it Swift}~J2058+05 \citep{Cenko2012}, {\it Swift}~J1112-8238 \citep{Brown2015}, and AT2022cmc \citep{Andreoni2022}.

The event rate of non-jetted TDEs is estimated to be $1\times10^{-5}\, {\rm yr^{-1}}$ per galaxy for a typical galaxy \citep{Donley2002, MT1999, WM2004}, whereas the event rate of jetted TDEs is estimated $0.03~{\mathrm{Gpc}}^{-3} {\mathrm{yr}}^{-1}$ \citep{Sun2015, 2020NewAR..8901538D, Andreoni2022}. Recent observations indicate that many {\it Changing look} active galactic nuclei (AGNs) have been associated with TDEs \citep{Eracleous1995, Merloni2015, Ricci2020, Ricci2021, Jana2021}. This could also potentially increase the population of TDEs with e-ROSITA \citep{2012Merloni} in the future. These event rates indicate the remarkable scarcity of the jetted TDE population. However, the population of jetted TDEs estimated above indicates a relatively small beaming angle ($\leq$ 1\textdegree). Of late, the discovery of off-axis jetted TDEs, such as IGR~J12580+0134 in the nucleus of NGC~4845 galaxy \citep{Walter2011} and the subsequent radio emission \citep{Lei2016} suggest that the TDEs, which emit hard X-rays with weak radio emission, can show a global property of X-ray TDEs. Several non-jetted, radio-emitting TDEs have also been detected \citep{alexander_radio_2020,cendes_radio_2021}. Moreover, the recently discovered association of astrophysical IceCube neutrinos with three TDE candidates \citep{Stein2021, Hayasaki2021} invoked a new era of multi-messenger astronomy of TDEs.

There are differences in X-ray spectral hardness between jetted TDEs and non-jetted TDEs. Non-jetted TDEs often exhibit softer spectra, where photon index $\Gamma > 2.5$ is associated with a thermal component \citep{Komossa2017}. Contrary to that, jetted TDEs have much harder spectra with $\Gamma \leq 2$. In addition, no clear evidence of a thermal component could be observed from the X-ray spectral energy distribution \citep{Magano2016, Seifina2017}. The domain of jetted TDEs began in March 2011 with the discovery of {\it Swift}~J1644+57. The origin of high-energy $\gamma$-rays is traced back to a galaxy at $z = 0.35$ \citep{Levan2011}. The peak X-ray luminosity exceeded the Eddington limit with the inferred black hole mass $3\times10^6 M_{\odot}$ \citep{Bloom2011b, Burrows2011, Levan2011}. The source was detected in radio \citep{Cendes2014, Berger2012, Zauderer2013}, infra-red (Wiersema et al. 2012); X-rays \citep{Reis2012, Saxton2012, GR2014, Magano2016, Zauderer2013}, $\gamma$-rays \citep{Bloom2011b, Burrows2011}. Unlike X-ray luminosity, which decayed over the next 500 days, the radio luminosity increased \citep{Berger2012, Zauderer2013} as the event progressed.

\begin{figure*}
\centering
\includegraphics[angle=0,width=8.5cm,height=6.5cm]{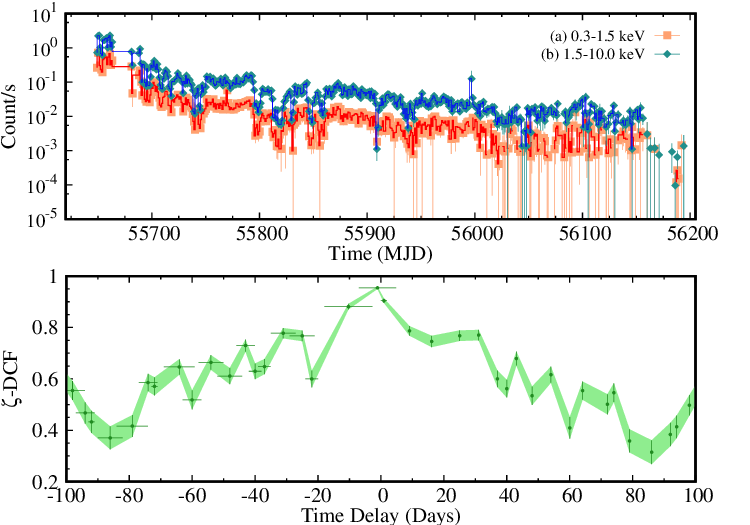}
\caption{{\it Upper Panel:} X-ray light curves of Swift~J1644+57, obtained from the {\it Swift}/XRT, are presented for 0.3--1.5 keV (square-light-salmon) and 1.5--10 keV energy bands (diamond-greenish-blu-e points). {\it Lower panel:} $\zeta$-DCF cross-correlation between 0.3-1.5 keV and 1.5--10 keV ranges are presented for Swift~J1644+57 using {\it Swift}/XRT data. Both the energy bands are highly correlated ($\rho_{max}=0.95$). Considering the error bars, no delay is observed between the two X-ray bands.}
\label{fig:lc}
\end{figure*}

\cite{Seifina2017} showed that the spectral hardness of the source evolved from $\Gamma \sim 1.8$ to $\Gamma \sim 1.2$ for the first 200 days since the outburst. However, the variation of spectral index met sudden fluctuations associated with the dips found in the light curve \citep{Magano2016}. From the X-ray spectral shape, luminosity, and the absence of the thermal component, \cite{Zauderer2013} asserted that the X-ray photons of {\it Swift}~J1644+57 could not have originated from the accretion disk. Rather, the origin of the X-rays could be from the internal dissipation within the jet at a large distance ($r \sim 10^4-10^5 r_{\rm g}$) from the SMBH. Contrary to their idea, \cite{Lei2016} suggested the X-ray emission from the off-axis jetted-TDE candidate IGR~J12580+0134 originates from the disk-corona. Beamed accretion, much like in the case of Blazers, could stand out as a possible scenario for the hyper accretion rate. However, in the case of Swift~J1644+57, the luminosity kept falling with decreasing X-ray photon index ($\Gamma$). This observed correlation between luminosity and $\Gamma$ opposes the ``harder when brighter” \citep{Krawczynski2004, Gliozzi2006, Zhang2006, PGW2017} nature of the X-ray bright Blazars and aligns with the accretion disk dominated Seyferts or galactic black holes (GBHs) \citep[e.g.,][]{Yang2015,AJ2022,AJ2026}. Stronger evidence of the accretion disk origin of the X-ray emission in \source~came from \cite{Kara2016}, where reverberation lag was observed.

A quasi-periodic oscillation (QPO) of 200~s, within the first 20 days from the peak luminosity, was discovered 
from the X-ray light curves in the 2-10 keV energy band, using the {\it Suzaku} and {\it XMM-Newton} observations \citep{Reis2012, Saxton2012}. Similar QPOs in X-rays have been observed for two other TDE candidates, namely 2XMM~J123103.2+110648 \citep{Lin2013} and ASASSN-14li \citep{Pasham2019}. \cite{NW2013} suggested the detection of a QPO within $0.0008$-$0.004$\, Hz for IGR J12580+0134, where the timescale is consistent for a TDE with central black hole mass $\sim 10^5 M_{\odot}$. There are many arguments that the QPOs can bridge the mass gap between the accretion processes of galactic, stellar-mass black holes, and SMBHs \citep{VU2005}. However, very few of them are detected with statistical confidence for AGNs \citep{Gierlinski2008, Alston2015, Ashton2021}. There are many challenges for observing QPOs originating from AGNs, apart from achieving a statistically significant {\it signal to noise ratio} \citep{VU2006}. Considering the mass of {\it Swift}~J1644+57, the 200~s QPO found in the X-ray light curve requires a scale of $\mathcal{O}(10)\,r_{\rm s}$ for a non-rotating SMBH. This scale fits within the range of the typical coronal radius parameter (hereafter $R_{cor}$) of TDEs as proposed by \cite{Mummery2021}, and is also consistent with the corona size, which is observationally suggested for many AGNs \citep{Alston2020}. Thus, we expect to find a reasonable explanation of hard X-ray variations through the disk-corona geometry by analyzing the existing data of jetted TDEs. The most promising mechanism for the luminous emission of TDEs is a thermal blackbody emission. \cite{Gezari2021} suggested that there is a bimodal distribution of the TDE thermal emission, indicating that the soft X-ray photons are produced from the inner disk region, whereas the optical/UV photons could be emitted from the photospheric outflows. The thermal photons produced from the disk can be a good candidate source for the Compton scattering. This suggests a disk-corona system can exist in X-ray TDEs.

\cite{Galeev1979} first proposed the disk-corona system to explain the hard X-ray origin of Galactic X-ray binaries. They formulated it based on an analogy with the solar coronal structure. The corona is heated through the magnetic reconnection at the disk surface and also through the convection within an accretion disk. As the inward flow progresses onto the black hole, the coronal structure can be dynamic, and the magnetic field strength increases, giving rise to a central cloud composed of hot electrons ($\sim 10^8$ K) and a magnetic field strength of the order of $10^7$ Gauss. This model can explain the hard X-ray emission from Cyg~X-1. \cite{HM1991} proposed a two-phase component model, where the standard accretion disk surrounding the corona emits the soft photons, which are also reprocessed in the corona via Comptonization and reflection. The model was able to explain the general features of AGNs spectra. Subsequently, \cite{Martocchia1996} and \cite{MF2004}  proposed the {\it Lamppost} model, which includes the effect of the gravitational bending due to the black hole spin and the reflection effect by the iron K$\alpha$ line \citep{Fabian1991, Tanaka1995}. Moreover, \cite{2002ApJ...572L.173L} constructed a simple theoretical disk-corona model, where the corona is heated by magnetic reconnection and cooled by Compton scattering. The origin of the very high-temperature corona was considered to be the magnetic field for all of the above-mentioned models. More recently, \cite{Done2012} provided a more observationally consistent model {\it Optxagnf}. This model provides a spectral fit originating from the corona and disk with sub to super-Eddington accretion flows of AGNs. Their model also provides a practical estimate for the coronal size. In the TDE context, \cite{Mummery2021} applied the simple empirical model, where the corona expands as the luminosity decays and the smallest coronal parameter ($R_{cor}$) corresponds to the peak luminosity, to model the hard X-ray profile seen in the six non-jetted TDEs. 
 
In this paper, we study how the Compton cloud of the jetted TDE {\it Swift}~J1644+57 varies with time by using the {\it Optxagnf} model and whether the model can explain the hard X-ray properties. The paper is structured in the following way. In Section~2, we explain the data analysis method and the subsequent processing. In Section~3, we describe our results obtained from data analysis. We then discuss our findings and the implications on TDEs in Section~4. Finally, we summarize our conclusions in Section~5.

%
\section{Data Analysis Process}
%
We used the archival data of {\it XMM-Newton} and {\it Swift}/XRT obtained through HEASARC\footnote{\url{http://heasarc.gsfc.nasa.gov/}}. We reprocessed all data using {\tt HEAsoft v6.26.1} \citep{Arnaud1996}, which includes {\tt XSPEC v12.10.1f}. We use the following cosmological parameters in this work: $H_0$ = 70 km s$^{-1}$ Mpc$^{-1}$, $\Lambda_0$ = 0.73, $\Omega_M$ = 0.27 \citep{Bennett2003}. With the cosmological parameters, the luminosity distance of \source~was considered 1402.8 Mpc \citep{Levan2011}.

\label{sec:data}
\subsection{Swift/XRT}
The X-ray telescope (XRT) \citep{Burrows2005} on board {\it Swift} has the highest cadence monitoring observation of Swift~J1644+57 in the X-ray band (0.3 to 10.0 keV), especially during the TDE from 2011. Depending on the brightness of the source, XRT observed this source in both Window Timing (WT) and Photon Counting (PC) modes. The source was observed by {\it Swift}/XRT over $\sim$500 times from 2011-03-31 to 2012-09-30. We stacked all observations into 17 segments depending on the brightness of the source. When the event started, the source was brighter. Thus, we stacked $\sim15$ observations into a single spectrum and produced the combined spectrum. This procedure is followed for the first $154$ observations. We stacked these observations into 10 combined observations (XRT1 to XRT10) (details are given in Table \ref{tab:log}). After that, the source became dimmed with respect to previous observations. For that, we stacked $\sim40$ observations into a single spectrum. This process is followed for the next 250 observations and produces the stacked spectrum from XRT11 to XRT15 (see Table \ref{tab:log}). After 2012-03-17, we stacked 176 observations and 108 observations into a combined observation for XRT16 and XRT17, respectively. We use the online tool ``XRT product builder''\footnote{\url{http://swift.ac.uk/user_objects/}} \citep{Evans2009} to extract the spectrum and light curves. This product builder performs all necessary processing and calibration and produces the final spectra and light curves of {\source~} WT and PC modes. The GRPPHA task is used with 10 counts per bin for XRT spectra. For more details, see Table \ref{tab:log}.

\subsection{XMM-Newton}
We reprocessed the {\it XMM-Newton} Epic-pn data using the Science Analysis Software (SAS) v17.0.0 and updated calibration files. We reprocessed the data with the ``epproc" task. For each observation ID, we created good time interval (GTI) files after eliminating the intervals of flaring particle background based on the count rate in the light curves above 10 keV. We filtered the event lists to get the good events with a pattern $<=$4 for a given GTI file and generated the cleaned event lists. We extracted source and background spectra using circular regions of radii 30 and 50 arcsec, respectively. We generated the light curves in $0.2-2$ keV and $3-10$ keV ranges with 10~s time bin and corrected from effects like vignetting, bad pixels, and variations in PSF. We also checked for any pile-up effects in the data. We generated the redistribution matrix and ancillary response files using the SAS tasks ``rmfgen" and ``afrgen", respectively. We grouped the spectrum with a minimum of 20 counts per bin and oversampled it by a factor of 3 using the ``specgroup" task.

	\begin{table*}
		\caption{Observation log for {\it Swift}/XRT and {\it XMM-Newton} observations of \source.}

		\label{tab:log}
		\begin{tabular}{lcccc}
			\hline
			ID&Date           & Obs. ID     & Instrument              & Total Exposure  \\
			&(yyyy-mm-dd)   &             &                          & (ks)     \\
			\hline
			&&&&\\
			XRT1 & 2011-03-31 & 00031955002 & {\it Swift}/XRT           &  224 \\
			&-2011-04-14 & -00031955016&                         &        \\
			XMM1 & 2011-03-31 & 0658400701 & {\it XMM-Newton}/Epic-pn & 26 \\

&&&\\
			XRT2 & 2011-04-15 & 00031955017 & {\it Swift}/XRT           &  210 \\
			&-2011-04-30 & -00031955032&                         &        \\
            XMM2 & 2011-04-16 & 0678380101 & {\it XMM-Newton}/Epic-pn & 25 \\
            XMM3 & 2011-04-30 & 0678380201 & {\it XMM-Newton}/Epic-pn & 29 \\
			XRT3 & 2011-05-01 & 00031955033 & {\it Swift}/XRT           &  137 \\
			&-2011-05-13 & -00031955047&                        &        \\
			XRT4 & 2011-05-14 & 00031955048 & {\it Swift}/XRT           &  162 \\
			&-2011-05-29 & -00031955063&                        &        \\
            XMM4 & 2011-05-16 & 0678380301 & {\it XMM-Newton}/Epic-pn & 30 \\
			XRT5 & 2011-05-30 & 00031955064 & {\it Swift}/XRT           &  125\\
			&-2011-06-12 & -00031955078&                        &        \\
            XMM5 & 2011-05-30 & 0678380401 & {\it XMM-Newton}/Epic-pn & 30 \\
			XRT6 & 2011-06-13 & 00031955079 & {\it Swift}/XRT           &  131 \\
			&-2011-06-26 & -00031955093&                        &        \\
			XRT7 & 2011-06-27 & 00031955094 & {\it Swift}/XRT           &  142 \\
			&-2011-07-13 & -00031955109&                        &        \\
            XMM6 & 2011-07-03 & 0678380501 & {\it XMM-Newton}/Epic-pn & 19 \\
			XRT8 & 2011-07-14 & 00031955110 & {\it Swift}/XRT           &  80 \\
			&-2011-07-28 & -00031955124&                        &        \\
            XMM7 & 2011-07-15 & 0678380601 & {\it XMM-Newton}/Epic-pn & 30 \\
            XMM8 & 2011-07-15 & 0678380701 & {\it XMM-Newton}/Epic-pn & 19 \\
			XRT9 & 2011-07-29 & 00031955125 & {\it Swift}/XRT           &  77 \\
			&-2011-08-12 & -00031955139&                        &        \\
			XRT10 & 2011-08-13 & 00031955140 & {\it Swift}/XRT           &  56 \\
			&-2011-08-27 & -00031955154&                        &        \\
            XMM9 & 2011-08-14 & 0678380801 & {\it XMM-Newton}/Epic-pn & 29 \\
            XMM10 & 2011-08-27 & 0678380901 & {\it XMM-Newton}/Epic-pn & 30 \\
			XRT11 & 2011-08-28 & 00031955155 & {\it Swift}/XRT           &  108 \\
			&-2011-10-01 & -00031955190&                        &        \\
            XMM11 & 2011-09-06 & 0678381001 & {\it XMM-Newton}/Epic-pn & 28 \\
            XMM12 & 2011-09-06 & 0678381101 & {\it XMM-Newton}/Epic-pn & 28 \\
            XMM13 & 2011-10-02 & 0678381201 & {\it XMM-Newton}/Epic-pn & 28 \\
			XRT12 & 2011-10-03 & 00031955191 & {\it Swift}/XRT           &  110 \\
			&-2011-11-11 & -00031955230&                        &        \\
            
			XRT13 & 2011-11-12 & 00031955231 & {\it Swift}/XRT           &  132 \\
			&-2011-12-21 & -00032200015&                        &        \\
           
			XRT14 & 2011-12-22 & 00032200016 & {\it Swift}/XRT           &  104 \\
			&-2012-01-30 & -00032200055&                        &        \\
			XRT15 & 2012-01-31 & 00032200056 & {\it Swift}/XRT           &  116 \\
			&-2012-03-17 & -00032200100&                        &        \\
			XRT16 & 2012-03-18 & 00032200101 & {\it Swift}/XRT           &  120 \\
			&-2012-06-07 & -00032200175&                        &        \\
			XRT17 & 2012-06-08 & 00032200101 & {\it Swift}/XRT           &  156 \\
			&-2012-09-30 & -00032526046&                        &        \\
			\hline
		\end{tabular}
	\end{table*}

\section{Results} 
\label{sec:results}

\subsection{Light curves} 
\begin{figure*}
\centering{
\includegraphics[angle=0,width=8.5cm,height=6.0cm]{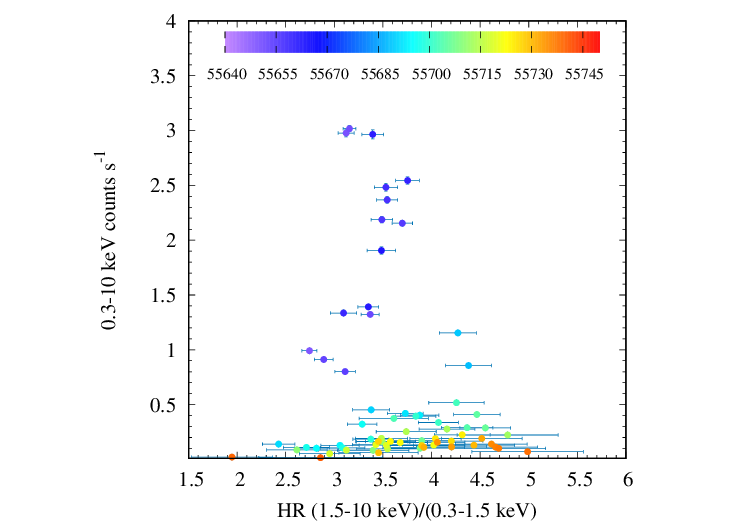}
\includegraphics[angle=0,width=8.5cm,height=6.0cm]{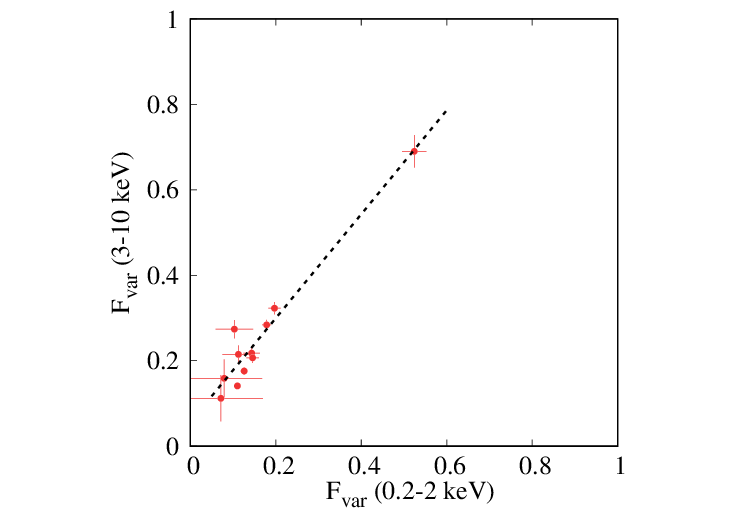}}
\caption{{\tt Left Panel:} Hardness ratio of Swift J1644+57 is presented. The source exhibited hard spectra throughout the tidal disruption event. The colorbar on top represents the MJD from the start of the event. {\tt Right Panel:} Correlated variability between 0.2--2 keV and 3--10 
keV is observed from \xmm data. The Pearson correlation coefficient is 0.96.}
\label{fig:HR}
\end{figure*}

The long-term X-ray light curve obtained from the \swift/XRT showed a high degree of variation during the TDE of \source. The source counts changed dramatically by orders of magnitude (see Fig.~\ref{fig:lc}) within the span of $\sim600$ days. We split the energy band of \swift into $0.3-1.5$ keV \& $1.5-10.0$ keV ranges. As evident from Fig.~\ref{fig:lc} and Fig.~\ref{fig:HR}, the hard band dominated the entire episode with sudden dips occasionally observed \citep{Magano2016}.

We performed cross-correlation analysis using $\zeta$-discrete cross-correlation function ({\tt ZDCF}\footnote{{\tt ZDCF: }\burl{http://www.weizmann.ac.il/particle/tal/research-activities/software}}, \citep{A97}) on the long-term light curve obtained from the \swift/XRT satellite. $\zeta$-DCF method provides a better estimate of the cross correlation function (CCF) for unevenly sampled data sets and considers the non-linearity while interpolating the lightcures. The cross-correlation study yields an optimal correlation of the 0.3-1.5 keV and 1.5-10 keV energy bands, having a correlation coefficient $\rho_{max}=0.95$. The correlation pattern peaked at a negative delay of $1.09\pm5.9$ days. Considering the measurement error, this could be considered as a zero-delay between the two energy bands of the X-ray. A similar zero-delay was also observed for an outbursting GBH, named XTE J1550-564, where a relatively smaller size of the accretion disk is attributed to the absence of any delay \citep{Chatterjee2020}. The light curve, along with the $\zeta$-discrete cross-correlation are presented in Fig.~\ref{fig:lc}. 

From the light curve of \swift/XRT, we obtained the Hardness-Ratio (HR) as presented in \cite{Magano2016} and \cite{Seifina2017}. We found that the source counts in the hard band were always greater (see Fig.~\ref{fig:HR}) than in the soft band. As the event progressed, the contribution of the hard band in the light curve also increased making harder spectra at the late time of the tidal disruption event. The left panel of Fig.~\ref{fig:HR} shows the `HR'-diagram obtained using the first 90 days of data. 

Using \xmm data, we calculated the fractional variability ($F_{\rm var}$) \citep{Edelson1996, Edelson2001, Edelson2012, Nandra1997, Vaughan2003} in the 0.2--2 and 3--10 keV bands. The light curves (presented as $x_i$ counts $s^{-1}$) contain uncertainties $\sigma_i$ of length $N$ with a mean $\mu$ and standard deviation $\sigma$, related by,

\begin{equation}
F_{\rm var} = \sqrt{\frac{\sigma^2_{\rm XS}}{\mu^2}},
\end{equation}

where, $\sigma^2_{\rm XS}$ is the excess variance \citep{Nandra1997, Edelson2002}.  $\sigma^2_{\rm XS}$ is defined as,

\begin{equation}
\sigma^2_{\rm XS}=\sigma^2 - \sigma^2_{\rm err},
\end{equation}

where, $\sigma^{2}_{\rm err}$ is the mean squared error. The $\sigma^{2}_{\rm err}$ is given by,

\begin{equation}
\sigma^2_{\rm err} =  \frac{1}{N}\sum_{i=1}^{N} \sigma^2_{\rm i}.
\end{equation}

The normalized excess variance is given by,

\begin{equation}
\sigma^2_{\rm NXS}=\frac{\sigma^2_{\rm XS}}{\mu^2}.
\end{equation}

The uncertainties in $F_{\rm var}$ and $\sigma_{\rm NXS}$ \citep{Vaughan2003} are given by,

\begin{equation}
{\rm err}(F_{\rm var})= \sqrt{\left( \sqrt{\frac{1}{2N}}\frac{\sigma_{\rm err}^2}{\mu^2 F_{\rm var}} \right)^2 + \left( \frac{1}{\mu}\sqrt{\frac{\sigma^2_{\rm err}}{N}} \right)^2},
\end{equation}

and

\begin{equation}
{\rm err}(\sigma^2_{\rm NXS}) = \sqrt{\left(\sqrt{\frac{2}{N}}\frac{\sigma_{\rm err}^2}{\mu^2} \right)^2 + \left(\sqrt{\frac{\sigma_{\rm err}^2}{N}}\frac{2F_{\rm var}}{\mu} \right)^2}.
\end{equation}

Table~\ref{tab:fvar} presents the results of our variability study in soft and hard bands. The excess variance or $\sigma^2_{NXS}$ yielded a negative value, which resulted in an imaginary fractional variability ($F_{var}$) for the XMM5 observation. The maximum of $\sigma^2_{NXS}$ or $F_{var}$ is observed during XMM3. The $F_{var}$ of the hard band is found to be higher than its softer counterpart for all XMM-Newton observations. We find the $F_{var}$ between 0.2-2 keV and 3-10 keV, presented in the right panel of Fig.~\ref{fig:HR}, are highly correlated with a correlation coefficient of 0.96.

\begin{table*}
\centering
\caption{Variability statistics in various energy ranges are shown in this table. We have opted for 100 sec time bins for variability analysis. In some cases, the average error of observational data exceeds the limit of $1\sigma$, resulting in negative excess variance. In such cases, we have imaginary $F_{var}$, which are not shown in the table.}
\begin{tabular}{lccccccc}
\hline
ID      & Energy band     &    $N$   &$x_{max}$ &$x_{min}$&$\frac{x_{max}}{x_{min}}$   &$\sigma^2_{NXS}$      &$F_{var}   $\\

&   keV           &       & Count/s  & Count/s &                 &$(10^{-2})     $      & $(\%)$\\
\hline
XMM1    &0.2-2.0          &    451    &8.34      &3.69     &2.26 & $1.21 \pm0.08$     & $11.0\pm0.51$ \\
XMM2    &0.2-2.0          &    378    &6.39      &1.77     &3.56 & $1.59 \pm0.16$     & $12.6\pm0.78$ \\
XMM3    &0.2-2.0          &    301    &1.46      &0.02     &60.8 & $27.4 \pm2.45$     & $52.4\pm2.83$ \\
XMM4    &0.2-2.0          &    540    &1.72      &0.37     &4.67 & $2.92 \pm3.19$     & $17.8\pm1.04$ \\
XMM5    &0.2-2.0          &    302    &0.30      &0.03     &11.8 &$-0.04 \pm2.95$     & $--$ \\
XMM6    &0.2-2.0          &    491    &1.52      &0.18     &8.60 & $3.90 \pm0.50$     & $19.7\pm1.41$ \\
XMM7    &0.2-2.0          &    322    &1.01      &0.13     &8.00 & $2.06 \pm0.52$     & $14.3\pm1.92$ \\
XMM8    &0.2-2.0          &    544    &0.94      &0.15     &6.16 & $2.12 \pm0.41$     & $15.6\pm1.46$ \\
XMM9    &0.2-2.0          &    513    &0.36      &0.03     &14.3 & $0.51 \pm0.14$     & $7.2\pm9.80$ \\
XMM10   &0.2-2.0          &    521    &0.49      &0.02     &20.1 & $1.27 \pm0.84$     & $11.2\pm3.75$ \\
XMM11   &0.2-2.0          &    386    &0.43      &0.02     &17.3 & $0.63 \pm1.41$     & $7.9\pm8.89$ \\
XMM12   &0.2-2.0          &    389    &0.77      &0.04     &17.8 & $1.12 \pm0.91$     & $10.3\pm4.41$ \\
\hline
XMM1    &3.0-10.0         &    451    &6.57      &2.25     &2.93 & $2.01 \pm0.12$     & $14.1\pm0.63$ \\
XMM2    &3.0-10.0         &    378    &5.28      &3.33     &2.92 & $3.10 \pm0.22$     & $17.6\pm0.89$ \\
XMM3    &3.0-10.0         &    300    &1.62      &0.02     &66.6 & $44.7 \pm3.53$     & $69.0\pm3.80$ \\
XMM4    &3.0-10.0         &    540    &2.04      &0.19     &10.8 & $8.15 \pm0.47$     & $28.4\pm1.20$ \\
XMM5    &3.0-10.0         &    299    &0.26      &0.03     &10.1 &$-0.03 \pm0.03$     & $--$ \\
XMM6    &3.0-10.0         &    491    &2.34      &0.20     &11.4 & $10.4 \pm0.63$     & $32.3\pm1.42$ \\
XMM7    &3.0-10.0         &    322    &1.36      &0.26     &5.31 & $4.81 \pm0.59$     & $21.8\pm1.60$ \\
XMM8    &3.0-10.0         &    544    &1.53      &0.23     &6.54 & $4.32 \pm0.41$     & $20.7\pm1.17$ \\
XMM9    &3.0-10.0         &    513    &0.36      &0.03     &14.0 & $1.30 \pm1.21$     & $11.2\pm5.41$ \\
XMM10   &3.0-10.0         &    521    &0.68      &0.03     &27.1 & $4.61 \pm0.86$     & $21.5\pm2.10$ \\
XMM11   &3.0-10.0         &    389    &0.44      &0.03     &17.3 & $2.54 \pm1.38$     & $15.9\pm4.40$ \\
XMM12   &3.0-10.0         &    387    &0.92      &0.05     &17.7 & $7.51 \pm1.05$     & $27.4\pm2.15$ \\
\hline
\end{tabular}
\label{tab:fvar}
\end{table*}

\subsection{Spectra}
To understand the X-ray spectral properties of \source, we used the phenomenological model along with the physical model. During fitting, we employed stacked XRT spectra to obtain a broad idea of spectral variations. All errors are quoted 
with 90\% confidence level, unless specified otherwise.

We started the spectral fitting with {\tt powerlaw} model, and the model in 
{\tt XSPEC} reads as: {\tt  TBabs*zTBabs*(powerlaw)}. 

\begin{figure*}
\centering {
\includegraphics[angle=0,width=0.66\columnwidth]{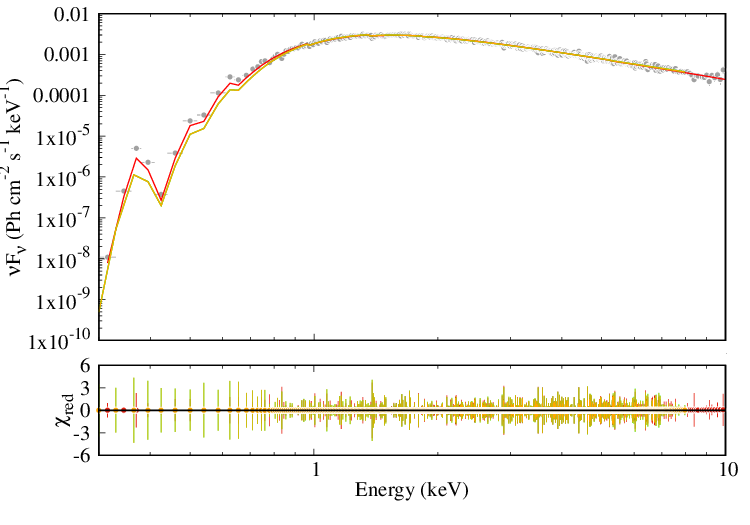}
\includegraphics[angle=0,width=0.66\columnwidth]{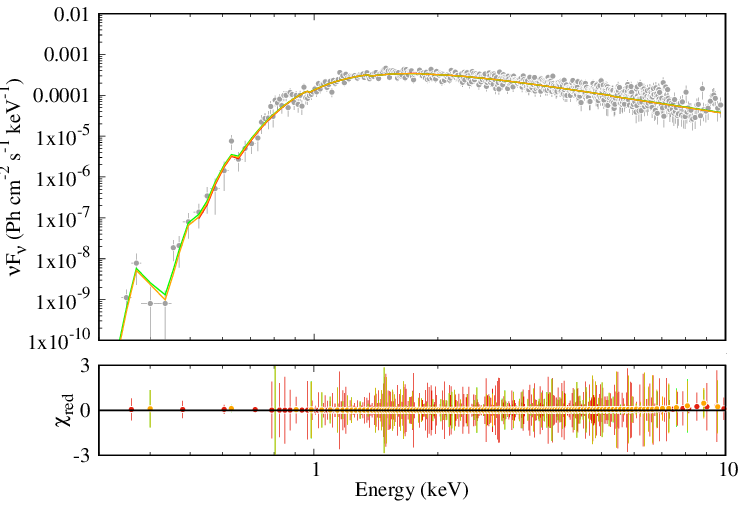}
\includegraphics[angle=0,width=0.66\columnwidth]{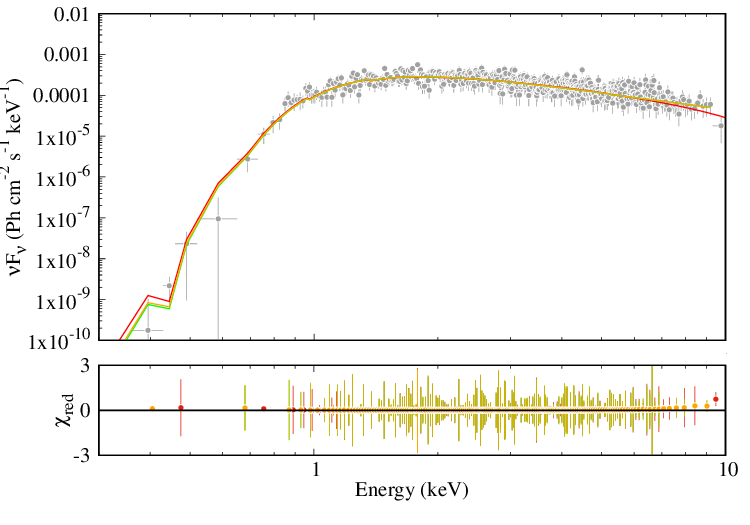}}
	\caption{{\tt left to right:} XRT2, XRT6, and XRT10 fitted unfolded spectra are presented for various models. The red line represents the {\tt Optxagnf}, green represents the {\tt Nthcomp}, and orange represents the \powerlaw model. Corresponding residues are plotted in the bottom panels. }
    
\label{fig:spec}
\end{figure*}

The galactic hydrogen column density ($N_H$) was kept frozen at $0.016\times10^{22}\mathrm{cm}^{-2}$ throughout the entire spectral fitting process. The initial fits with the \powerlaw model provide a good fit with $\chi^2_{red} \sim 1$. Table~\ref{tab:powerlaw} presents the findings of the fits. During the first observation, which spans from MJD 55651 to MJD 55665, the spectral index of the source was observed to be $1.79\pm0.02$. Corresponding $N_H$ was found to be $1.75\pm0.04\times10^{22}\mathrm{cm}^{-2}$. The normalization of the {\tt powerlaw} component was observed to be the highest $0.0177\pm0.005$ \phc among all other observations. Our second observation was made by stacking \swift/XRT spectra within the duration of MJD 55666 to MJD 55681. The spectrum was fitted with a power law index $\Gamma=1.62\pm0.01$ and corresponding $N_H$ was found to be $1.44\pm0.03\times10^{22}\mathrm{cm}^{-2}$. The normalization and flux, presented in Table \ref{tab:powerlaw}, were found to be decreased compared to the first observation. This is consistent with the light curve. We found the highest reduced chi-square ($\chi^2_{red}\sim1.21$) for this particular observation. The fitted unfolded spectrum is presented in the left panel of Fig. \ref{fig:spec}.
The following observation considered the stacked spectra within MJD 55682 to MJD 55694. The source flux decreased substantially in the XRT3 observation. We found a higher $N_H$ compared to the previous observation. The spectral indices had increased in XRT4 and XRT5 compared to the previous observations, XRT2 and XRT3. Apart from that, the $N_H$ had increased, and the corresponding flux decreased. We found a general trend of decreasing $\Gamma$ up to XRT11. The corresponding $N_H$ also increased, and the luminosity exhibited a decreasing pattern. The findings were comparable with the findings of \cite{Seifina2017}. However, from XRT12, the spectral index started to increase. However, it should be noted that the errors corresponding to the spectral indices became larger as the source became dim and data quality was reduced. For XRT16 observation, we found the source spectrum showed a steeper $\Gamma$ at $1.72\pm0.22$, comparable to the initial days of the TDEs. The left Panel of Fig. \ref{fig:contour} shows the $90\%$ confidence contours of $N_H$ and $\Gamma$. The spectral parameter variations, fitted with the \powerlaw model, are presented in Fig.~\ref{fig:powerlaw}.

\xmm observed \source~twelve times within the duration of the initial phase of the tidal disruption event ranging from MJD 55667 to MJD 55836. We fitted the \xmm data with the \powerlaw model to compare the parameter variations with the stacked observations created using \swift data. During the first observation, referred to as XMM1, we found that the source spectrum had a spectral index of $1.69\pm0.01$ with a relatively lower hydrogen column density of $1.29\pm0.01\times10^{22}\mathrm{cm}^{-2}$. The \powerlaw normalization and flux are found to be highest (see Table~\ref{tab:powerlaw}) in that particular observation. In the following observation, XMM2, the spectrum hardened with $\Gamma=1.54\pm0.01$ and associated $N_H$ increased to $1.42\pm0.02\times10^{22}\mathrm{cm}^{-2}$. The normalization and the flux were reduced substantially. Later, on MJD 55697, we fitted the XMM3 spectrum where $\Gamma$ was found highest among all the \xmm observations, having a value of $1.87\pm0.05$. The flux decreased dramatically to $-11.14\times10^{-11}$ from $-10.15\times10^{-11}$ \ecs. This sudden dip in the flux could be associated with the dips found in the long-term light curve \citep{Magano2016, Seifina2017}. During XMM4 observation, the source spectrum returned to a harder spectrum with $\Gamma=1.55\pm0.02$. The concurrent flux increased from the previous observation. Later, in the XMM5 observation, we found an upward variation of the spectral index with $\Gamma=1.59\pm0.08$. However, considering the error, the spectral index could be in a similar range as the XMM4 observation. The normalization and flux decreased rapidly, and the observed values are the lowest among all the \xmm observations. Thereafter, both the flux and normalization increased with decreased spectral index. Post XMM5, the \powerlaw fitted spectra exhibited a steady decrease in the flux with occasional variation in the spectral index. The source normalization also varied from $13.56\times10^{-3}$ to $0.78\times10^{-3}$ \phc throughout the \xmm observations.

The results obtained from the baseline \powerlaw model, presented in Table~\ref{tab:powerlaw}, suggest a good fit using a single power law component. We found no residuals around the low energy part, which could indicate the presence of soft excess \citep{Singh1985, Arnaud1985} or thermal component from the disk, often visible for X-ray TDE \citep{Gezari2021}. The \swift and \xmm data within the overlapping time broadly exhibited similar spectral variation with increasing $N_H$ associated with decreasing spectral index. The overall flux of \source~decreased. We found a lower and more confined variation of $N_H$ from the \xmm observations as compared to the \swift observations. However, the variation of $N_H$ from the observations with both satellites remained similar throughout the entire period. We did not find the presence of the iron K$\alpha$ line throughout the disruption event. The degrees of freedom for \powerlaw fitting were higher for \swift observations compared to the \xmm observations. 

\subsubsection{Optxagnf}
The {\tt Optxagnf} model, proposed by \cite{Done2012}, provides an in-depth understanding of the spectral parameters as the parameters are generated from a physical picture of the accretion disk around the AGNs. The model-fitted parameters include the luminosity, coronal radius parameter ($R_{cor}$), spin of the central black hole, soft-excess temperature, optical depth, spectral index, and the fraction of Comptonized photons. It should be noted that {\tt Optxagnf} is the only publicly available model to fit the X-ray spectra of AGNs where $R_{cor}$  could be estimated. We employed this model to fit the XRT spectra. The model in {\tt xspec} reads as: {\tt  TBabs*zTBabs*(optxagnf)}.

 The model spectrum computes both soft-excess and Comptonized photons. The accretion rate governed by the mass of the central black hole determines the luminosity of the source (in units of Eddington luminosity - $log(L/L_{Edd})$). The geometry of the {\tt Optxagnf}~ model separates the accretion disk into three regions. First, the disk emission is produced through a color temperature-corrected blackbody emission from radii $R_{\rm out} > r > R_{\rm cor}$, where $R_{\rm out}$ and $R_{\rm cor}$ are the outer edge of the disk and the corona, respectively. Luminosity is computed keeping the distance and mass constant, while the mass-to-energy conversion efficiency factor ($\eta$) depends on the spin. Second, at $r<R_{\rm cor}$, the emission is dominated by Comptonized emission from a warm and optically thick medium. The third region where inverse Comptonization occurs is the optically thin corona around the disk, and the fraction of inverse Comptonization can be estimated from the $f_{pl}$ parameter during the spectral fitting. The temperature and optical depth of the warm, optically thick medium could be found from $kT_s$ and $\tau$, respectively. The model calculates the primary emission of Comptonized photons using the {\tt Nthcomp} model, where the seed photon temperature is kept fixed at $r=R_{\rm cor}$, and the electron temperature of the optically thin hot corona is frozen at 100 keV. We freeze the mass of the black hole at $3\times10^6~M_{\odot}$ \citep{Bloom2011b}. The Galactic $N_H$ along the line of \source~ remained frozen at $0.016\times10^{22}\mathrm{cm}^{-2}$ while performing the fitting. According to the model, it was suggested to keep the normalization frozen at unity. The redshift $z=0.354$ and distance $d=1402.8$ Mpc are also kept fixed \citep{Levan2011}. The $log(r_{out})$ is kept fixed at 5 since the apogee of extremely elongated TDEs, capable of explaining lengthy disruption events \citep{Hayasaki2013, Dai2015, Hayasaki2018}, could reach up to $10^4-10^5\,r_{\rm g}$. The remaining parameters are left free to vary. It should be noted that the reported luminosities in Table \ref{tab:optx} are isotropic and model-dependent without employing the beaming corrections. From Table~\ref{tab:powerlaw}, we can see that the power law model fits the spectra of \source~ well, and the soft X-ray excess is absent in the spectra. Thus, we freeze the $kT_s$ at 0.1 keV and $\tau$ at a maximum value of 5.0. Initially, we left the $f_{pl}$ thawed. We observed that the $f_{pl}$ values saturated in most cases. Later, better fits were obtained while fitting the spectra with variable $kT_s$ and $\tau$ and is presented in Table \ref{tab:optx}. We observed that the warm corona temperature and optical depth remained high across all observations except XRT3. We found that the $kT_s$ gradually increases as the event progresses and remains within $0.59\leq kT_s \leq 5.73$. On the other hand, the variation of optical depth remained non-monotonic within the limit $1.25 \leq \tau \leq 7.83$. The variation of $f_{pl}$ remained within the range of $0.48\leq f_{pl} \leq0.93$ during our observation period. For TDEs, this is expected and is conjectured by \cite{Chakraborty2026}. {\tt Optxagnf} fitted spectra for three epochs are presented in Fig. \ref{fig:spec}, and the variation of fitted parameters with time is plotted in the 
{\tt right panel} of Fig. \ref{fig:powerlaw} and are presented in the Table \ref{tab:optx}.

\begin{figure*}
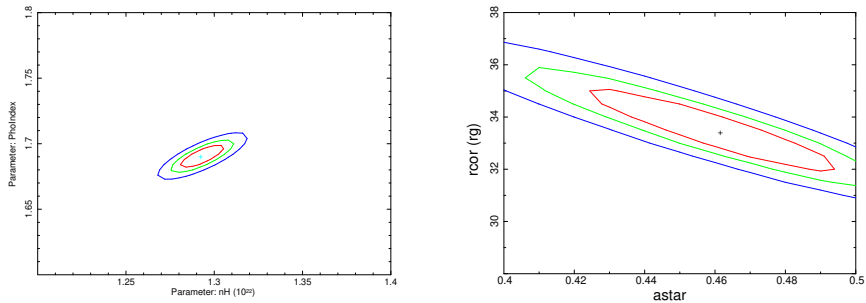

	\centering {
		\includegraphics[width=5.5cm,height=6.5cm,angle=270]{con_power.eps}
		\includegraphics[width=5.5cm,height=6.5cm,angle=270]{optxagnf.eps}}
\caption{Contours of confidence are plotted for \swift/XRT observations fitted with \powerlaw and {\it Optxagnf} model.}
\label{fig:contour}
\end{figure*}

We fitted the  0.3--10 keV stacked \swift spectra for the entire episode of TDE. The spectra fitted alone with the {\tt Optxagnf} model provided good statistics with $\chi^2_{red}\sim1$. The model-fitted spectral parameters are presented in Table \ref{tab:optx}. We found the luminosity of the source exceeded nearly 500 times the Eddington luminosity during the initial few days of the event. This agrees with the earlier studies like \cite{Zauderer2013} and is also seen from the current fitting with {\tt Optxagnf}~model. Over time, the luminosity of the source decreased nearly 200 times from the initial phase, but never went into the sub-Eddington regime during the period of our observation. The variation of $N_H$ remained similar to what has been observed during the fitting with the \powerlaw model. The spin of the central black hole remained within the range of $0.4<a^*<0.8$, making it an intermediate spinning black hole. During the initial days, the coronal parameter of the source was observed at $31\pm2~r_{\rm g}$. The right panel of Fig. \ref{fig:contour} represents the confidence contours of spin ($a$) vs $R_{cor}$. The coronal radius parameter was found to vary within $31<R_{cor}<65~r_{\rm g}$. We only used the stacked \swift data for {\tt Optxagnf} fitting as it provides better data quality to understand the long-timescale variations of \source~during its tidal disruption episode. 

\begin{table*}
\caption{ {\tt powerlaw} fitting result for the 0.3--10 keV energy range.}
\begin{center}
\begin{tabular}{c c c c c c c}
			\hline
			ID    & MJD      & $N_H$                 &$\Gamma$               & N                   & log(Flux)                     & $\chi^2/dof$   \\
			      &  (days)        & $(10^{22}~cm^{-2})$    &                       & $(10^{-3})$         & ($10^{-11}$~erg/cm$^2$/s)      &                \\
			\hline
			&&&\swift/XRT&&&\\
			\hline
			XRT1  &$55658$   &$1.75^{+0.04}_{-0.04}$ &$1.79^{+0.02}_{-0.02}$ &$17.7^{+5.50}_{-5.50}$ & $-09.94$  & $902.74/827$   \\
			XRT2  &$55673.5$ &$1.44^{+0.03}_{-0.03}$ &$1.62^{+0.01}_{-0.01}$ &$10.8^{+2.12}_{-2.12}$ & $-10.09$  & $1047.47/861$  \\
			XRT3  &$55688$   &$1.67^{+0.08}_{-0.08}$ &$1.63^{+0.04}_{-0.04}$ &$4.51^{+2.74}_{-2.74}$ & $-10.48$  & $741.23/866$   \\
			XRT4  &$55702.5$ &$1.90^{+0.08}_{-0.08}$ &$1.75^{+0.04}_{-0.04}$ &$3.10^{+1.73}_{-1.73}$ & $-10.63$  & $525.19/703$   \\
			XRT5  &$55717.5$ &$1.99^{+0.10}_{-0.10}$ &$1.74^{+0.05}_{-0.05}$ &$2.20^{+1.60}_{-1.60}$ & $-10.83$  & $574.73/643$   \\
			XRT6  &$55731.5$ &$1.95^{+0.11}_{-0.11}$ &$1.61^{+0.05}_{-0.05}$ &$1.47^{+1.13}_{-1.13}$ & $-10.96$  & $615.79/640$   \\
			XRT7  &$55747$   &$2.13^{+0.15}_{-0.15}$ &$1.62^{+0.07}_{-0.07}$ &$0.96^{+0.96}_{-0.96}$ & $-11.14$  & $596.64/594$   \\
			XRT8  &$55763$   &$1.89^{+0.13}_{-0.13}$ &$1.45^{+0.07}_{-0.07}$ &$1.26^{+0.12}_{-0.12}$ & $-10.96$  & $555.14/596$   \\
			XRT9  &$55778$   &$1.94^{+0.16}_{-0.16}$ &$1.56^{+0.07}_{-0.07}$ &$1.33^{+0.14}_{-0.14}$ & $-10.98$  & $508.38/544$   \\
			XRT10 &$55793$   &$2.21^{+0.20}_{-0.20}$ &$1.40^{+0.09}_{-0.09}$ &$1.12^{+0.15}_{-0.15}$ & $-10.98$  & $498.03/520$   \\
			XRT11 &$55818$   &$2.04^{+0.22}_{-0.22}$ &$1.44^{+0.10}_{-0.10}$ &$0.46^{+0.07}_{-0.07}$ & $-11.39$  & $431.81/475$   \\
			XRT12 &$55856.5$ &$2.14^{+0.19}_{-0.19}$ &$1.57^{+0.10}_{-0.10}$ &$0.72^{+0.09}_{-0.09}$ & $-11.25$  & $466.99/509$   \\
			XRT13 &$55896.5$ &$2.15^{+0.19}_{-0.19}$ &$1.50^{+0.09}_{-0.09}$ &$0.58^{+0.07}_{-0.07}$ & $-11.31$  & $441.98/519$   \\
			XRT14 &$55936  $ &$2.18^{+0.39}_{-0.39}$ &$1.61^{+0.16}_{-0.16}$ &$0.38^{+0.10}_{-0.10}$ & $-11.54$  & $274.05/309$   \\
			XRT15 &$55980  $ &$1.94^{+0.24}_{-0.24}$ &$1.55^{+0.12}_{-0.12}$ &$0.32^{+0.06}_{-0.06}$ & $-11.59$  & $337.79/371$   \\
			XRT16 &$56044.5$ &$2.23^{+0.55}_{-0.55}$ &$1.72^{+0.22}_{-0.22}$ &$0.21^{+0.08}_{-0.08}$ & $-11.84$  & $166.80/197$   \\
			XRT17 &$56143  $ &$1.76^{+0.80}_{-0.80}$ &$1.28^{+0.34}_{-0.34}$ &$0.04^{+0.02}_{-0.02}$ & $-12.40$  & $99.32/104$    \\
			&&&&&&\\
			\hline
			&&&XMM-Newton&&&\\
			\hline
			&&&&&&\\
			XMM1 &$55667  $ &$1.29^{+0.01}_{-0.01}$ &$1.69^{+0.01}_{-0.01}$ &$13.56^{+0.10}_{-0.10}$ & $-10.02$  & $224.69/174$    \\
			XMM2 &$55681  $ &$1.42^{+0.02}_{-0.02}$ &$1.54^{+0.01}_{-0.01}$ &$ 8.87^{+0.10}_{-0.10}$ & $-10.15$  & $186.68/173$    \\
			XMM3 &$55697  $ &$1.33^{+0.06}_{-0.06}$ &$1.87^{+0.05}_{-0.05}$ &$ 1.18^{+0.07}_{-0.08}$ & $-11.14$  & $141.20/133$    \\
			XMM4 &$55711  $ &$1.49^{+0.04}_{-0.03}$ &$1.55^{+0.02}_{-0.02}$ &$ 2.48^{+0.07}_{-0.07}$ & $-10.71$  & $193.00/167$    \\
			XMM5 &$55745  $ &$1.41^{+0.14}_{-0.13}$ &$1.59^{+0.08}_{-0.08}$ &$ 0.27^{+0.03}_{-0.03}$ & $-11.68$  & $97.00/90  $    \\
			XMM6 &$55757  $ &$1.61^{+0.05}_{-0.05}$ &$1.39^{+0.02}_{-0.02}$ &$ 1.68^{+0.06}_{-0.06}$ & $-10.80$  & $186.57/162$    \\
			XMM7 &$55769  $ &$1.62^{+0.06}_{-0.06}$ &$1.41^{+0.03}_{-0.03}$ &$ 1.44^{+0.07}_{-0.06}$ & $-10.87$  & $190.57/151$    \\
			XMM8 &$55787  $ &$1.63^{+0.05}_{-0.05}$ &$1.34^{+0.02}_{-0.02}$ &$ 1.41^{+0.05}_{-0.05}$ & $-10.85$  & $179.59/163$    \\
			XMM9 &$55800  $ &$1.43^{+0.10}_{-0.10}$ &$1.31^{+0.05}_{-0.05}$ &$ 0.30^{+0.02}_{-0.02}$ & $-11.51$  & $107.40/129$    \\
			XMM10&$55810  $ &$1.58^{+0.08}_{-0.08}$ &$1.39^{+0.04}_{-0.04}$ &$ 0.58^{+0.03}_{-0.03}$ & $-11.26$  & $154.57/145$    \\
			XMM11&$55824  $ &$1.51^{+0.11}_{-0.12}$ &$1.41^{+0.06}_{-0.06}$ &$ 0.37^{+0.03}_{-0.03}$ & $-11.46$  & $154.39/121$    \\
			XMM12&$55836  $ &$1.51^{+0.08}_{-0.08}$ &$1.34^{+0.04}_{-0.04}$ &$ 0.78^{+0.04}_{-0.05}$ & $-11.10$  & $167.89/144$    \\

\hline	
\end{tabular}
\end{center}
\label{tab:powerlaw}
\end{table*}

\section{Discussion}
%
We studied \source~during its tidal disruption event of 2011-2012 using \swift and \xmm data within the energy range 0.3-10 keV. We employed the \powerlaw as our phenomenological model and a physical model, such as {\tt Optxagnf}, to understand the behavior of the source as the event progresses. Being a jetted-TDE, the source exhibited variability in both the spectral and temporal domains and is one of the most studied tidal disruption events to date. During our analysis, we found that there are several similarities between TDEs and the declining phase of an outbursting GBH. The similarities between GBH transients and TDEs are seen from their spectral energy distributions and the decaying luminosity with time. The X-ray spectral indices of GBHs are found to decrease in the luminosity decay phase \citep{MR2009}. Likewise, the decreasing tendency of the spectral index has also been observed in {\it Swift} J1644+57 \citep{Seifina2017}. Recent {\it NICER} observations \citep{Gendreau2012} of MAXI J1820+070 provided evidence of a dynamically contracting corona for the onset stage of the flare \citep{Kara2019}. \cite{Jana2016} explained, using the two-component advective flow model \citep{CT1995}, why the Compton cloud expanded at the declining phase of a low inclination GBH, MAXI J1836-194, during the 2011 outburst. The variation of the corona size is also observed for AGN IRAS 13224-3809 \citep{Alston2020}. Apart from $R_{cor}$ variation, relativistic jet launching is another phenomenon that GBH transients and TDEs have in common \citep{MR1994, Patra2019, Espi2020, Blandford2019, Zauderer2011}. 

\subsection{Nuclear Properties}
\begin{figure*}
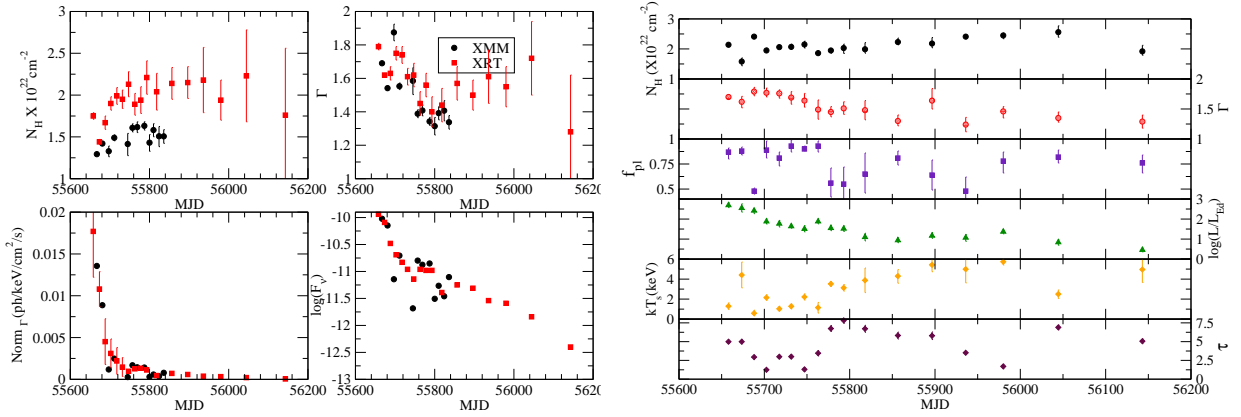

\centering {
\includegraphics[angle=0,width=0.97\columnwidth]{xw_xmm_pow_comp.eps}
\includegraphics[angle=0,width=0.97\columnwidth]{opt_param_time_2026.eps}}
\caption{Variation of spectral parameters of Swift~J1644+57 are presented for {\tt powerlaw} and 
{\tt optxagnf} model. Only stacked {\it Swift}/XRT data are employed to fit {\tt optxagnf} model.}
\label{fig:powerlaw}
\end{figure*}

X-ray emission from \source~has been observed for around 600 days. The hard X-ray component was found to dominate, and the radiation is conjectured to be beamed \citep{Zauderer2013} and many authors. From our spectral analysis, we find that the source exhibited variations in hydrogen column density ($N_H$). Fig.~\ref{fig:powerlaw}, represents the variation of \powerlaw fitted parameters. We find that the hydrogen column density, at the beginning, was higher at $1.75\pm0.04\times10^{22}\mathrm{cm}^{-2}$. Later, in XRT2, the $N_H$ dropped to $1.44\pm0.04\times10^{22}\mathrm{cm}^{-2}$. Post XRT2, the spectra showed a gradual increase in the value of $N_H$ and plateaued at a late stage. Similar observational findings are also seen through \xmm spectra fitted with \powerlaw model and \swift spectra fitted with the {\tt Optxagnf} model. The variation of $N_H$ could be a generic property of the jetted-TDEs.  

\cite{Seifina2017} reported that the spectral index of individual spectra observed by \swift/XRT saturates after MJD 55730; i.e., after eighty days since the start of the event. We used stacked spectra of \swift/XRT where the spectral indices are observed to be decreasing up to MJD 55800. However, after that, the spectral indices increased. But, finally, at XRT17 observation, the spectral index was $1.28\pm0.34$. Both \swift and \xmm data are employed to measure the spectral indices variation within MJD 55650 to MJD 55800, where a similar pattern has been found. Using the {\tt Optxagnf} model, we have also observed similar variations. The increase in the X-ray spectral indices after 150 days could be associated with the increased optical activity after $\sim 30-50$ days since the initial observations \citep{Levan2016}. We find that $N_H$ and $\Gamma$ are weakly anti-correlated with a Pearson Correlation coefficient of -0.22.

The {\tt Optxagnf} model provides several parameters of the source. Among them, we find that the luminosity, in terms of $\log(L/L_{Edd})$, is highly sensitive to the fitting. The fitted spectra exhibited a variation of the order of 2 with the range of our observations. During the initial days, \source~had a bolometric luminosity of around 500 times the Eddington limit. In the entire observation, the source luminosity was in the super-Eddington regime. The luminosity followed a steady decrease from the initial days, with occasional variation in between, which can be associated with the intrinsic photon count variation in the light curve. During XRT17, the last observation, the source luminosity decreased to nearly 3 times the Eddington limit. We find the source luminosity weakly correlated with the hydrogen column density, with a Pearson Correlation coefficient of -0.31.

The spin of \source, measured using the $a^*$ parameter of {\tt Optxagnf} model, exhibited a region of an intermediate spinning black hole. We find $a^*$ to vary from 0.4 to 0.8. The value of the spin aligns with the earlier claim by \cite{Tch2014}, where they predicted the spin of the source should be greater than $\geq 0.5$.

%
\subsection{Coronal Variations}
%
The hard X-ray photons, emerging from the vicinity of a black hole, are thought to originate from a corona which can be formed through magnetic reconnection in the accretion disk \citep{Galeev1979,2002ApJ...572L.173L}. There are a few practical models of corona that have been proposed so far: {\it Lampost} model \citep{Martocchia1996, MF2004} provides physical explanations of the X-ray spectra from the AGNs; \cite{Done2012} constructs {\tt Optxagnf} model,  computes the size of the corona, $R_{cor}$ , from the spectral fitting. Also, it has been observed that the nature of the corona is dynamic for GBHs \citep{Kara2019} and for the AGNs \citep{Alston2020}. That is why it is crucial to explore how the corona size evolves with time for TDEs.

\begin{figure*}
\centering
\includegraphics[angle=0,width=1.15\columnwidth, height = 0.75\columnwidth]{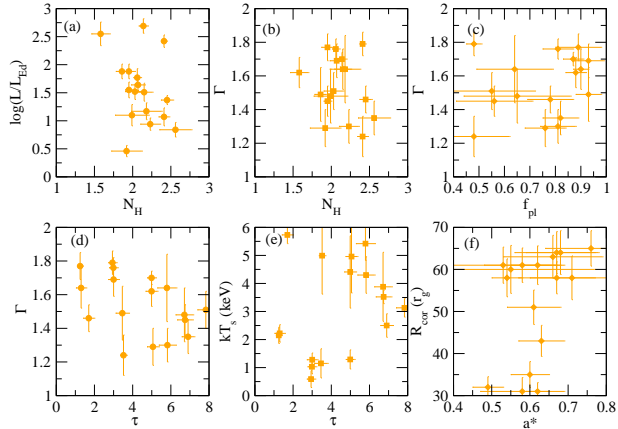}
\caption{{\tt Optxagnf} fitted spectral parameters variation with each other of Swift~J1644+57 are presented. Panel (a) shows the variation of $N_H$ with respect to the normalized luminosity. The variation of $N_H$ with respect to $\Gamma$ is shown in panel (b). Panel (c) demonstrates the weak correlation between $\Gamma$ and scattering fraction $f_{pl}$. Variation of the $\Gamma$ and $\tau$ can be observed in panel (d). Panel (e) presents the variation of 
$R_{cor}$ and spin parameter. Uncorrelated variation between spin and $R_{cor}$ can be observed in 
panel (e).}
\label{fig:optxagnf_corr}
\end{figure*}

To measure the $R_{cor}$ variation of \source, we employ the {\tt Optxagnf} model to determine the spectral fits using the stacked \swift/XRT data. Through our spectral fitting, we find that at the peak luminosity of \source, the radial size of the corona was $31\pm2.3~r_{\rm g}$. Subsequently, an almost similar size of the $R_{cor}$ was found in the XRT2 and XRT3 observational epochs. The size of the corona increases $35\pm3.6~r_{\rm g}$ in the XRT4 epoch. Later on, in the XRT5 and XRT6 epochs, the $R_{cor}$  keeps increasing to $43\pm3.7~r_{\rm g}$ and $51\pm4.3~r_{\rm g}$, respectively. During the XRT7 observation, the $R_{cor}$  is $58\pm4.1~r_{\rm g}$. Up to the XRT7 observation, we observed a sharp rise in the $R_{cor}$ . The fitted value of the $R_{cor}$  is $61\pm4.2~r_{\rm g}$ at XRT8 and drops to $58\pm4.5~r_{\rm g}$ at XRT10. From the XRT10 to XRT17 observations, the $R_{cor}$  varies little from $58\,r_{\rm g}$ to $64\,r_{\rm g}$, indicating a saturation level of the parameter. All estimated values of the coronal parameters are depicted in Fig.~\ref{fig:rcor}. In addition, we find through our analysis that the $R_{cor}$  is anti-correlated with the luminosity $\log(L/L_{\rm Edd})$ with the Pearson correlation coefficient: $-0.87$.

Recently, \cite{Mummery2021} proposed the disk-corona model to explain non-thermal hard-X-ray emissions from X-ray TDEs. According to their prescription, the size of the corona would depend on the disk luminosity as follows:
\begin{equation}
R_{cor} = \begin{cases*}
R_{I},                              & initial size of corona,\\
R_I + (R_C-R_I)f(l),              & intermediate size of corona,\\
R_C,                              & final size of corona,
\end{cases*}
\label{eq:2}
\end{equation}
where 
\begin{eqnarray}
f(l)=\frac{l_I-l(t)}{l_I-l_{C}},
\nonumber
\end{eqnarray}
$l_I$, $l_{C}$, and $l(t)$ are the initial, final, and the time-dependent Eddington ratios, respectively. Starting with an initial $R_{cor}$ $R_{I}$, the corona expands following the decay pattern of the luminosity profile and finally saturates to a value of $R_{C}$. We adopt $R_I=30\,r_{\rm g}$ and $R_C=65\,r_{\rm g}$. The normalized disk luminosity, $\log(L/L_{\rm Edd})$, at each observational epoch is displayed in Table~\ref{tab:optx}. The predicted $R_{cor}$ at each epoch using Equation~(\ref{eq:2}) is denoted with red circles in Fig.~\ref{fig:rcor}. The error in the theoretical values is a contribution arising from the observational error in the measurement of luminosity as presented in Table~\ref{tab:optx}. We find that the evolution of theoretically speculated corona size agrees with that of the observational $R_{cor}$  parameter in \source, and considering the error bars, their variation remains similar.

Fig.~\ref{fig:cartoon} gives a schematic view of the disk-corona-jet system for the early and late stages of \source. The physical scale of the disk and corona fits within the range of $2R_{\rm t}\sim100\,r_{\rm g}$ for a $10^6~M_{\odot}$ SMBH, where $R_{\rm t}$ is given by Equation~(\ref{eq:1}). We presume that, in the initial days, the accretion rate was at the super-Eddington rate, and the corona was small. As the days progressed, the accretion rate reduced, and the magnetic flux became stronger, as suggested by \cite{Tch2014}. This allows the disk to have a bigger corona than that of the initial state. Although the disk is yet in the super-Eddington regime at this moment, the bigger corona is likely to produce a larger number of high-energy X-ray photons via inverse Compton scattering. Within the first hundred days, the jet flux increases monotonically and is anti-correlated with the X-ray flux \citep{Berger2012, Seifina2017}, suggesting that the jet region also expands during this period. Depending on the quality and amount of the X-ray observational data, one can measure the coronal variation of the other three jetted TDEs and non-jetted TDEs using the same analysis as ours. This is also useful for improving the theoretical corona model in TDEs. More recently, \cite{Chakraborty2026} argued that the accretion disk of X-ray bright TDEs could be composed of a variable corona having lower electron temperature and higher optical depth, and can also be observed from the variation of parameters presented 
in Table \ref{tab:optx}. 

\begin{figure*}
\centering {
\includegraphics[angle=0,width=0.97\columnwidth]{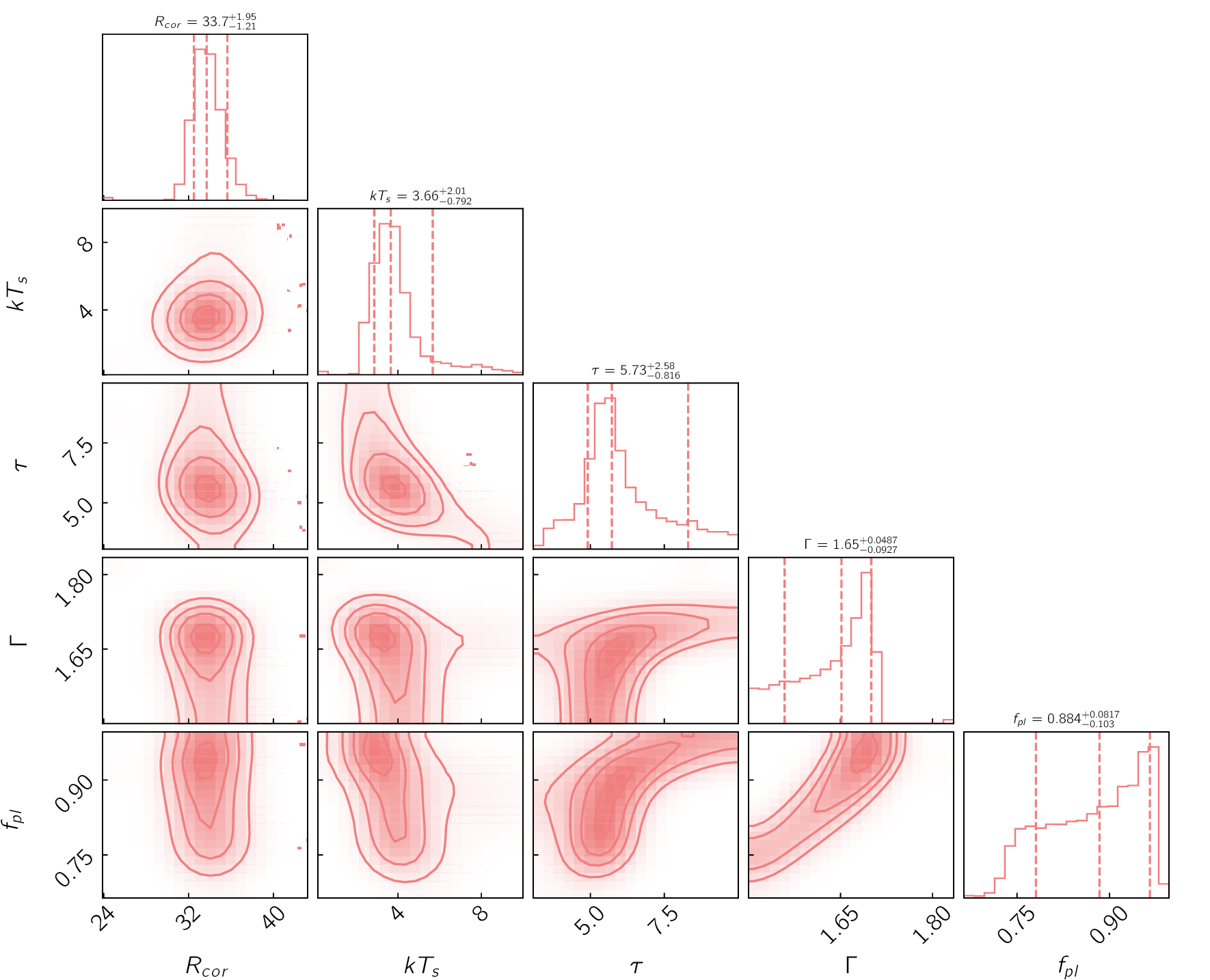}
\includegraphics[angle=0,width=0.97\columnwidth]{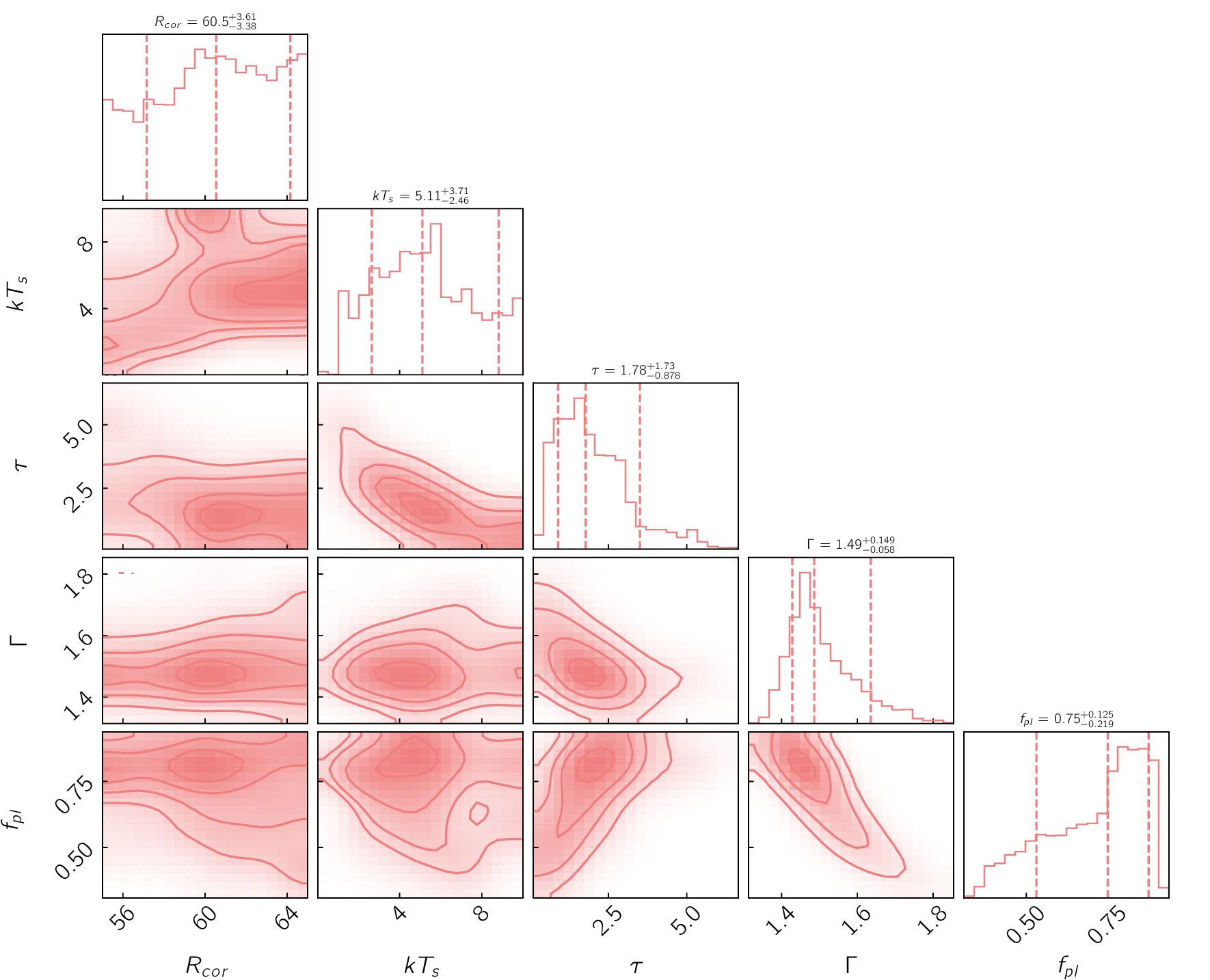}}
	\caption{Markov Chain Monte-Carlo plot for parameter variations of XRT2 and XRT15 are presented.
    Goodman-Weare (EMCEE) \citep{FM2013} is opted for 32 walkers and a chain length of 1 million 
    iterations, where the transient phase is removed by burning the chain up to 9984 iterations. The 
    spin parameter was kept frozen at $a^*=0.60$ to perform the MCMCs.}
\label{fig:mcmc}
\end{figure*}

\begin{figure*}
\centering
\includegraphics[angle=0,width=1.0\columnwidth]{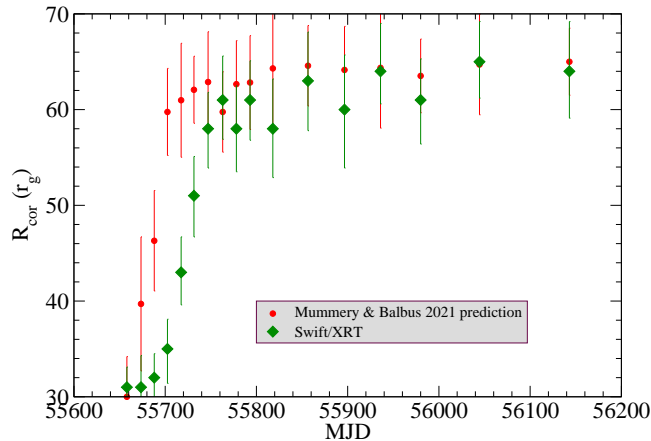}
	\caption{Coronal variation with time is presented for Swift~J1644+57. \swift/XRT fitted values are presented 
    by green-diamond points. The red circles denote the theoretical values of $R_{cor}$  as predicted by \cite{Mummery2021}.}
\label{fig:rcor}
\end{figure*}

\begin{figure*}
\centering
\includegraphics[angle=0,width=13.5cm,height=6.5cm]{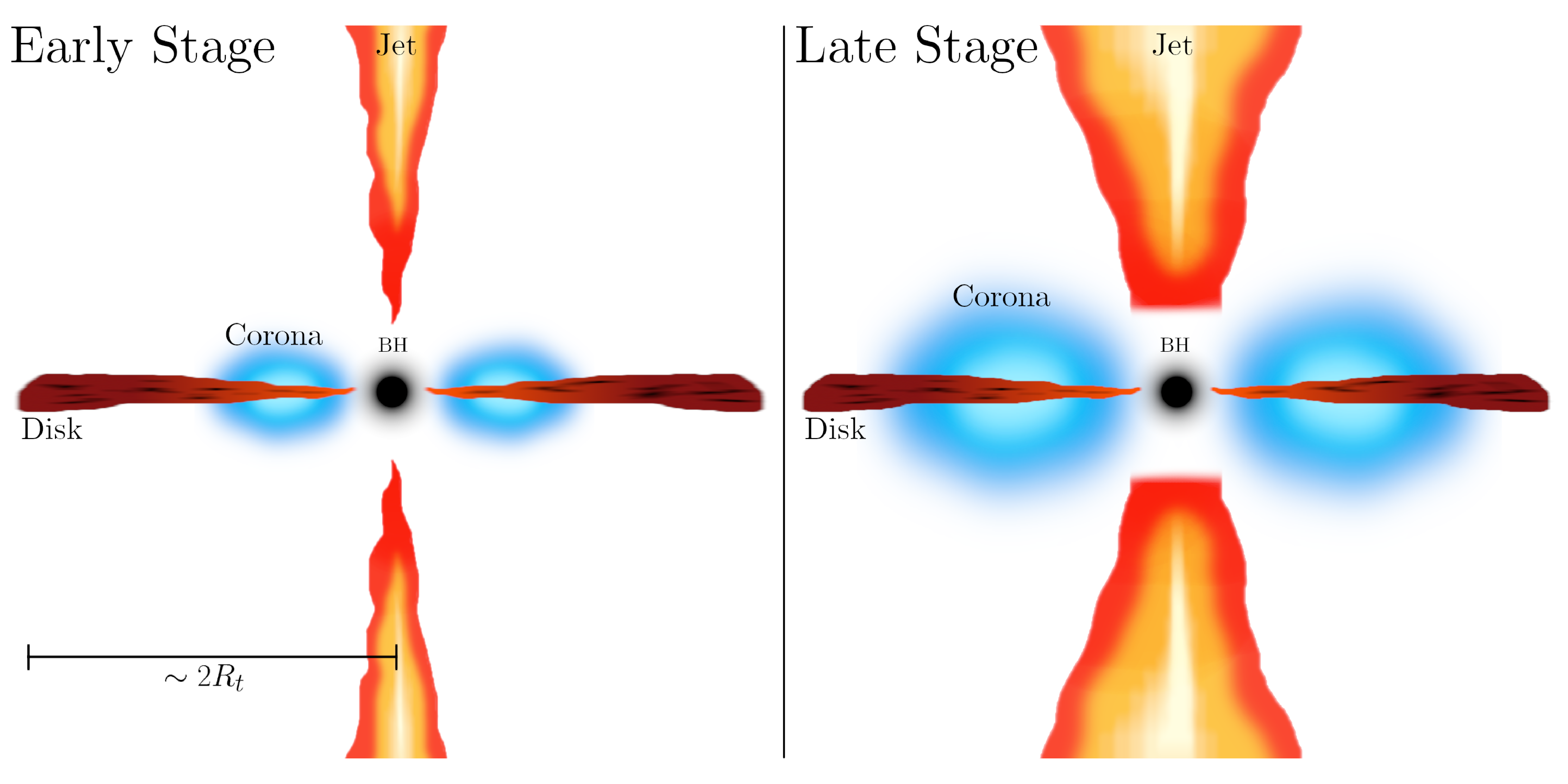}
\caption{Edge-on view schematic diagram of $R_{cor}$ variation in the early and late stages of \source. The system comprises a disk around the SMBH, a corona on the disk, and a jet. The disk size is twice the tidal disruption radii, which is approximately estimated to be $2R_{\rm t}\sim100\,r_{\rm g}$ for the $10^6M_\odot$ SMBH and a solar-type star (see equation~{\ref{eq:1}}).}
\label{fig:cartoon}
\end{figure*}

\subsubsection{Viability of TDE Corona Models}
It should be noted that, to date, there exists no publicly available model in XSPEC to 
fit the hyper accretion regime of Swift J1644+57. The source was one of the very few 
jetted TDEs ever observed with a QPO feature. The initial accretion rate was $\sim1000~L_{Edd}$ 
and was widely explored for its unique attributes. While choosing the model, one must be 
extremely cautious. {\tt Optxagnf} is the only model in XSPEC that applies to the 
super-Eddington accretion regime of the AGNs, and measures the coronal radius parameter. 

According to the {\tt Optxagnf} model \citep{Done2012}, in the $r\leq r_{cor}$ region, 
the disk and the corona could co-exist to make a ``disk-corona” structure 
\citep{Galeev1979, HM1991, Martocchia1996, 2002ApJ...572L.173L}, and such a 
configuration also provides higher photon interception to the corona. The transition 
radius after which the radiation transfers from pure color-corrected black body 
to the inverse Comptonizations in the high $\tau$ \& low $kT_s$ and/or low $\tau$ \& high $kT_s$ 
regions is called the coronal radius. In reality, Compton up or down scattering in the 
radiation pressure supported torus is complex in nature \citep{Kawaguchi2003} due to 
the interplay between twisted magnetic field lines \citep{JSD2019} and the velocity field of 
electrons \citep{AC2018}. The Compton y-parameter is inadequate for interpreting 
Comptonized spectra in super‑Eddington, jet‑dominated accretion flows, as it assumes 
a thermal, homogeneous, isotropic electron plasma in the Thomson regime. In high‑$\tau$, 
advective, or non‑thermal environments such as Swift J1644+57, y could become degenerate 
and physically misleading. Nevertheless, X-ray emission originated 
from the ``disk-corona”, or corona was earlier suggested by \cite{Lei2016, Kara2016, Magano2016}. 
Moreover, the source exhibits Comptonized X-ray spectra with variable hardness that are 
well described by evolving Comptonization components, independent of the 
precise inner accretion flow geometry \citep{Bloom2011a, Burrows2011}.

Recently, \cite{Chakraborty2026} has proposed a coronal emission for TDEs (see their Fig. 7) 
where the optical depth could be higher, accompanied by a lower electron temperature. 
The parameters obtained from the {\tt Optxagnf} model, presented in Table \ref{tab:optx}, exhibit 
a similar nature of corona. Although the morphology of the TDE corona could depart from the 
structure of the warm corona, the thermal properties, such as high $\tau$ and $kT_s$, resemble 
the warm corona of the {\tt Optxagnf} model. The variation of $R_{cor}$ could be observed 
from the MCMC plots presented in the Fig. \ref{fig:mcmc}. However, it should be noted that 
{\tt Optxagnf}, developed for AGNs, may not be physically self-consistent for hyper-Eddington 
jetted TDEs, such as for \source.

The high optical depth ($\tau$) \& low ($kT_s$) corona could form in the case of  
radiation pressure supported tori or super-Eddington thick disks, which were investigated 
by \cite{Abramowicz1978, Kozlowski1978, ACN1980, PW1980, Sikora1981, Fukue1982, SKC1985, 
Madau1988, Ulmer1997, JSD2019, JD2024}. Thick disks are capable of launching collimated 
jets \citep{RBBP1982, SKC1985b} and could also produce QPOs through oscillations \citep{Jaroszynski1986}.
The radiative properties of the TDEs need to be explored in this regime. However, there are 
no models in XSPEC with the thick disk feature. Furthermore, under the ideal situation, one 
could be exploring the Magnetically Arrested Disks (MAD, \cite{NIA2003, Tch2014}) to fit the 
broadband SED spectra of the jetted TDEs. In future work, we will incorporate a simplified 
thick disk model in {\tt XSPEC} to examine the parameters for several X-ray bright 
TDEs.

\begin{table*}
\centering
\caption{{\tt Optxagnf} fitted spectral parameter are presented. We used stacked {\it Swift/}XRT over {\it XMM-Newton} as the former provides more degrees of freedom over the latter. The following parameters are kept frozen while fitting: $N_H^{Gal} = 0.016\times10^{22}\mathrm{cm}^{-2}$, model normalization $N_{optx}=1$, redshift $z=0.354$, distance $d=1402.8$ Mpc, and $log(r_{out})=5$.}
\begin{tabular}{lcccccccccc}
	\hline
       ID&      MJD     &    $N_H$           &$\log(\frac{L}{L_{Edd}})$ &$a^*$ & $R_{cor}$    &$kT_s$      &$\tau$ & $\Gamma$ &$f_{pl}$ & $dof/chi^2$\\
	 &    (days)    &($10^{22}cm^{-2}$)         &                        &     &  ($r_g$)   &(keV)       &       &          &         &            \\
	\hline
		&&&&&&&&&&\\
        XRT1&55658&$2.14^{+0.06}_{-0.07}$&$2.69^{+0.13}_{-0.12}$&$0.62^{+0.03}_{-0.04}$&$31^{+2.1}_{-2.3}$&$1.29^{+0.34}_{-0.33}$&$4.99^{+0.23}_{-0.24}$&$1.70^{+0.04}_{-0.04}$&$0.87^{+0.04}_{-0.07}$&$887/830$\\
        &&&&&&&&&&\\
		XRT2&55673.5&$1.58^{+0.12}_{-0.14}$&$2.55^{+0.21}_{-0.20}$&$0.58^{+0.11}_{-0.13}$&$31^{+3.3}_{-4.1}$&$4.41^{+1.29}_{-1.28}$&$4.99^{+0.31}_{-0.27}$&$1.62^{+0.09}_{-0.10}$&$0.88^{+0.04}_{-0.04}$&$1058/938$\\
		&&&&&&&&&&\\
		XRT3&55688&$2.41^{+0.04}_{-0.03}$&$2.42^{+0.11}_{-0.15}$&$0.49^{+0.04}_{-0.06}$&$32^{+2.5}_{-2.2}$&$0.59^{+0.30}_{-0.25}$&$2.94^{+0.22}_{-0.22}$&$1.79^{+0.07}_{-0.07}$&$0.48^{+0.03}_{-0.03}$&$722/689$\\
		&&&&&&&&&&\\
		XRT4&55702.5&$1.95^{+0.08}_{-0.07}$&$1.88^{+0.13}_{-0.13}$&$0.60^{+0.05}_{-0.04}$&$35^{+3.1}_{-3.6}$&$2.15^{+0.29}_{-0.26}$&$1.25^{+0.15}_{-0.15}$&$1.77^{+0.08}_{-0.07}$&$0.89^{+0.09}_{-0.08}$&$625/703$\\
	    &&&&&&&&&&\\
	    XRT5&55717.5&$2.06^{+0.05}_{-0.05}$&$1.77^{+0.16}_{-0.17}$&$0.63^{+0.06}_{-0.06}$&$43^{+3.7}_{-3.4}$&$1.03^{+0.25}_{-0.22}$&$2.99^{+0.23}_{-0.25}$&$1.76^{+0.06}_{-0.08}$&$0.81^{+0.06}_{-0.08}$&$575/646$\\
	    &&&&&&&&&&\\
	    XRT6&55731.5&$2.07^{+0.09}_{-0.08}$&$1.64^{+0.10}_{-0.10}$&$0.61^{+0.07}_{-0.06}$&$51^{+4.1}_{-4.3}$&$1.28^{+0.24}_{-0.26}$&$3.00^{+0.28}_{-0.28}$&$1.69^{+0.10}_{-0.11}$&$0.93^{+0.06}_{-0.06}$&$683/638$\\
	    &&&&&&&&&&\\
	    XRT7&55747&$2.15^{+0.11}_{-0.12}$&$1.51^{+0.16}_{-0.15}$&$0.67^{+0.10}_{-0.10}$&$58^{+3.8}_{-4.1}$&$2.22^{+0.31}_{-0.32}$&$1.30^{+0.27}_{-0.27}$&$1.64^{+0.12}_{-0.11}$&$0.90^{+0.03}_{-0.02}$&$593/587$\\
	    &&&&&&&&&&\\
	    XRT8&55763&$1.86^{+0.08}_{-0.06}$&$1.88^{+0.13}_{-0.12}$&$0.62^{+0.07}_{-0.08}$&$61^{+4.6}_{-4.1}$&$1.15^{+0.52}_{-0.54}$&$3.46^{+0.35}_{-0.36}$&$1.49^{+0.16}_{-0.16}$&$0.93^{+0.05}_{-0.06}$&$550/594$\\
	    &&&&&&&&&&\\
	    XRT9&55778&$1.95^{+0.05}_{-0.09}$&$1.55^{+0.14}_{-0.13}$&$0.54^{+0.04}_{-0.04}$&$58^{+4.5}_{-4.5}$&$3.52^{+0.24}_{-0.24}$&$6.74^{+0.41}_{-0.38}$&$1.45^{+0.09}_{-0.08}$&$0.56^{+0.15}_{-0.14}$&$510/547$\\
	    &&&&&&&&&&\\
	    XRT10&55793&$2.03^{+0.13}_{-0.18}$&$1.52^{+0.14}_{-0.14}$&$0.58^{+0.10}_{-0.10}$&$61^{+4.1}_{-4.2}$&$3.13^{+0.33}_{-0.33}$&$7.83^{+0.44}_{-0.39}$&$1.51^{+0.11}_{-0.10}$&$0.55^{+0.17}_{-0.17}$&$511/527$\\
	    &&&&&&&&&&\\
	    XRT11&55818&$1.99^{+0.22}_{-0.13}$&$1.10^{+0.19}_{-0.19}$&$0.71^{+0.11}_{-0.15}$&$58^{+5.2}_{-5.1}$&$3.88^{+1.23}_{-1.23}$&$6.71^{+0.46}_{-0.46}$&$1.48^{+0.16}_{-0.16}$&$0.65^{+0.21}_{-0.19}$&$426/463$\\
	    &&&&&&&&&&\\
	    XRT12&55856.5&$2.23^{+0.13}_{-0.09}$&$0.94^{+0.11}_{-0.12}$&$0.66^{+0.13}_{-0.09}$&$63^{+5.1}_{-5.2}$&$4.30^{+0.65}_{-0.71}$&$5.82^{+0.45}_{-0.47}$&$1.30^{+0.10}_{-0.08}$&$0.81^{+0.07}_{-0.07}$&$444/461$\\
	    &&&&&&&&&&\\
	    XRT13&55896.5&$2.18^{+0.20}_{-0.14}$&$1.17^{+0.15}_{-0.13}$&$0.55^{+0.22}_{-0.16}$&$60^{+5.7}_{-6.1}$&$5.42^{+0.60}_{-0.68}$&$5.79^{+0.50}_{-0.51}$&$1.64^{+0.20}_{-0.14}$&$0.64^{+0.15}_{-0.15}$&$440/508$\\
	    &&&&&&&&&&\\
	    XRT14&55936&$2.41^{+0.07}_{-0.07}$&$1.07^{+0.16}_{-0.18}$&$0.67^{+0.11}_{-0.12}$&$64^{+5.0}_{-3.4}$&$4.99^{+1.37}_{-1.37}$&$3.52^{+0.17}_{-0.17}$&$1.24^{+0.12}_{-0.12}$&$0.48^{+0.14}_{-0.13}$&$246/300$\\
	    &&&&&&&&&&\\
	    XRT15&55980&$2.45^{+0.08}_{-0.11}$&$1.37^{+0.08}_{-0.11}$&$0.53^{+0.14}_{-0.10}$&$61^{+4.3}_{-4.6}$&$5.73^{+0.31}_{-0.33}$&$1.70^{+0.29}_{-0.31}$&$1.46^{+0.08}_{-0.11}$&$0.78^{+0.09}_{-0.09}$&$329/368$\\
	    &&&&&&&&&&\\
	    XRT16&56044.5&$2.56^{+0.21}_{-0.17}$&$0.84^{+0.13}_{-0.15}$&$0.76^{+0.12}_{-0.10}$&$65^{+4.2}_{-3.8}$&$2.50^{+0.42}_{-0.44}$&$6.90^{+0.33}_{-0.33}$&$1.35^{+0.10}_{-0.07}$&$0.82^{+0.07}_{-0.06}$&$166/192$\\
	    &&&&&&&&&&\\
	    XRT17&56143&$1.92^{+0.20}_{-0.13}$&$0.46^{+0.10}_{-0.10}$&$0.68^{+0.09}_{-0.08}$&$64^{+5.2}_{-4.9}$&$4.96^{+1.30}_{-1.30}$&$5.06^{+0.31}_{-0.32}$&$1.29^{+0.11}_{-0.11}$&$0.76^{+0.08}_{-0.10}$&$99.5/100$\\
	    &&&&&&&&&&\\

\hline
\end{tabular}
\label{tab:optx}
\end{table*}


%
\section{Summary}
%
We have re-visited the X-ray emission of the jetted-TDE \source~between 2011 and 2012 by analyzing the archival \swift and \xmm data with HEAsoft. \source~showed a massive tidal disruption event during 2011 and 2012, where the flux of the hard (3--10\,keV) photons dominated that of the soft (0.2--2\,keV) photons over the entire period. We have hypothesized that those hard photons originate from the Compton cloud, i.e., the corona, and have estimated the size of the corona by measuring the size of the hard X-ray emitted region using the publicly available model {\tt Optxagnf}. Our key findings are summarized as follows.

\begin{enumerate}
\item
 
We find that the spectral index non-monotonically decreases with time, demonstrating that the fraction of harder photons increases with time. This indicates that the size of the Compton cloud gets larger because the soft photons are more scattered to be exhausted. We observed that the optical depth ($\tau$) and the temperature of the
warm corona ($kT_s$) remained high -- a result that aligns with the  \cite{Chakraborty2026} study.

\item 

By interpreting the hard X-ray-emitted region as a corona, we have estimated the size of the corona as a function of time. Considering the {\tt Optxagnf} model, we confirm that the coronal radius parameter $R_{cor}$ increases over time for the observed period. We also find that the evolution of the corona size agrees with the theoretical conjecture made by \cite{Mummery2021}. 

\item
We find that the flux of hard and soft photons is highly correlated with the correlation coefficient $\rho_{max}=0.95$ throughout the event. Their correlation pattern has peaked at a zero delay. Moreover, the flux variabilities of the soft (0.2-2 keV) and hard (3-10 keV) photons correlate with each other with a Pearson correlation coefficient of $0.96$. These correlations suggest that the soft and hard band photons are emitted from the same source.

\end{enumerate}

Four jetted TDEs have been observed so far \citep{Cenko2012, Brown2015, Andreoni2022}.
In the future, more exotic events could be observed by \textit{XRISM} \citep{Tashiro18}. The 
large field of view and imaging capability of the proposed-to-NASA mission \textit{AXIS} 
\citep{Mushotzky2019, Reynolds2023, SSH2023, KOss2025} could provide a detailed structure of the 
X-ray emitting region of these sources. Also, to perform the short-term timing studies and search 
for more QPOs among the AGNs, the large effective area and high throughput of the proposed-to-CSA  
\textit{Colibr\`i} \citep{Heyl2019, Caiazzo2019} and proposed-to-ESA \textit{NewAthena} 
\citep{Newathena2025} missions will be required in the future. Future proposed X-ray interferometric missions 
\citep{Cash2000, Cash2003, GC2004, AC2017, AC2018, AC_review_2018, Uttley2020, Gandhi2023, Weaver2026} 
are essential to visualize the coronal variation in the configuration space, much like what EHT \citep{EHT2025} 
has obtained in the radio and sub-mm range.


%
%
\newpage
\begin{center}
\section*{acknowledgments}
\end{center}
The authors acknowledge an anonymous reviewer for their constructive comments.
AC acknowledges the UPES Seed grant (UPES/R\&D/SoAE/25062025/23) for partially supporting this research. 
The research of A.C., S.H., and S.S.H. were supported by the Canadian Space Agency (CSA) and the Natural Sciences and Engineering Research Council of Canada (NSERC) through the Discovery Grants and the Canada Research Chairs programs. The Basic Science Research Program has supported K.H. through the National Research Foundation of Korea (NRF), funded by the Ministry of Education (2016R1A5A1013277 (K.H. and A.C.), 2020R1A2C1007219, and RS-2025-23323627 (K.H.)). 
K.H. thanks Andrew Mummery for fruitful discussions. AJ acknowledges support from the Fondecyt fellowship (Proyecto 3230303). NK acknowledge funds from European Union - Next Generation EU, Mission 4 Component 1 CUP C53D23001330006 and INAF Large Grant 2023 BLOSSOM F.O. 1.05.23.01.13.
Research work at the Physical Research Laboratory, Ahmedabad, is funded by the Department of Space, Government of India. This research has made use of data and/or software provided by the High Energy Astrophysics Science Archive Research Center (HEASARC), which is a service of the Astrophysics Science Division at NASA/GSFC and the High Energy Astrophysics Division of the Smithsonian Astrophysical Observatory. This research has made use of the NASA/IPAC Extragalactic Database (NED), which is operated by the Jet Propulsion Laboratory, California Institute of Technology, under contract with the National Aeronautics and Space Administration. This research has made use of the SIMBAD database, operated at CDS, Strasbourg, France. 



\bibliography{ref-TDE}{}

@ARTICLE{Dai2015,
       author = {{Dai}, Lixin and {McKinney}, Jonathan C. and {Miller}, M. Coleman},
        title = "{Soft X-Ray Temperature Tidal Disruption Events from Stars on Deep Plunging Orbits}",
      journal = {\apjl},
         year = 2015,
        month = oct,
       volume = {812},
       number = {2},
          eid = {L39},
        pages = {L39},
          doi = {10.1088/2041-8205/812/2/L39},
archivePrefix = {arXiv},
       eprint = {1507.04333},
 primaryClass = {astro-ph.HE},
       adsurl = {https://ui.adsabs.harvard.edu/abs/2015ApJ...812L..39D}
}

@ARTICLE{Yang2015,
       author = {{Yang}, Qi-Xiang and {Xie}, Fu-Guo and {Yuan}, Feng and {Zdziarski}, Andrzej A. and {Gierli{\'n}ski}, Marek and {Ho}, Luis C. and {Yu}, Zhaolong},
        title = "{Correlation between the photon index and X-ray luminosity of black hole X-ray binaries and active galactic nuclei: observations and interpretation}",
      journal = {\mnras},
         year = 2015,
        month = feb,
       volume = {447},
       number = {2},
        pages = {1692-1704},
          doi = {10.1093/mnras/stu2571},
archivePrefix = {arXiv},
       eprint = {1412.1358},
 primaryClass = {astro-ph.HE},
       adsurl = {https://ui.adsabs.harvard.edu/abs/2015MNRAS.447.1692Y}
}

@ARTICLE{AJ2026,
       author = {{Jana}, Arghajit and {Ricci}, Claudio and {Tortosa}, Alessia and {Dimopoulos}, George and {Trakhtenbrot}, Benny and {Bauer}, Franz E. and {Temple}, Matthew J. and {Koss}, Michael and {Gupta}, Kriti Kamal and {Chang}, Hsian-Kuang and {Diaz}, Yaherlyn and {Illic}, Dragana and {Kallov{\'a}}, Krist{\'\i}na and {Shablovinskaya}, Elena},
        title = "{Multiwavelength properties of changing-state active galactic nuclei: I. The evolution of soft excess and X-ray continuum}",
      journal = {\aap},
         year = 2026,
        month = mar,
       volume = {707},
          eid = {A213},
        pages = {A213},
          doi = {10.1051/0004-6361/202556654},
archivePrefix = {arXiv},
       eprint = {2601.07337},
 primaryClass = {astro-ph.HE},
       adsurl = {https://ui.adsabs.harvard.edu/abs/2026A&A...707A.213J}
}

@ARTICLE{AJ2022,
       author = {{Jana}, Arghajit},
        title = "{Global accretion properties of black hole X-ray binaries: A phenomenological perspective}",
      journal = {\mnras},
         year = 2022,
        month = dec,
       volume = {517},
       number = {3},
        pages = {3588-3597},
          doi = {10.1093/mnras/stac2939},
archivePrefix = {arXiv},
       eprint = {2210.05283},
 primaryClass = {astro-ph.HE},
       adsurl = {https://ui.adsabs.harvard.edu/abs/2022MNRAS.517.3588J}
}

@ARTICLE{PGW2017,
       author = {{Pandey}, Ashwani and {Gupta}, Alok C. and {Wiita}, Paul J.},
        title = "{X-Ray Intraday Variability of Five TeV Blazars with NuSTAR}",
      journal = {\apj},
         year = 2017,
        month = jun,
       volume = {841},
       number = {2},
          eid = {123},
        pages = {123},
          doi = {10.3847/1538-4357/aa705e},
archivePrefix = {arXiv},
       eprint = {1705.02719},
 primaryClass = {astro-ph.HE},
       adsurl = {https://ui.adsabs.harvard.edu/abs/2017ApJ...841..123P}
}

@ARTICLE{Zhang2006,
       author = {{Zhang}, Y.~H. and {Treves}, A. and {Maraschi}, L. and {Bai}, J.~M. and {Liu}, F.~K.},
        title = "{XMM-Newton View of PKS 2155-304: Hardness Ratio and Cross-Correlation Analysis of EPIC pn Observations}",
      journal = {\apj},
         year = 2006,
        month = feb,
       volume = {637},
       number = {2},
        pages = {699-710},
          doi = {10.1086/498498},
archivePrefix = {arXiv},
       eprint = {astro-ph/0510040},
 primaryClass = {astro-ph},
       adsurl = {https://ui.adsabs.harvard.edu/abs/2006ApJ...637..699Z}
}

@ARTICLE{Gliozzi2006,
       author = {{Gliozzi}, M. and {Sambruna}, R.~M. and {Jung}, I. and {Krawczynski}, H. and {Horan}, D. and {Tavecchio}, F.},
        title = "{Long-Term X-Ray and TeV Variability of Mrk 501}",
      journal = {\apj},
         year = 2006,
        month = jul,
       volume = {646},
       number = {1},
        pages = {61-75},
          doi = {10.1086/504700},
archivePrefix = {arXiv},
       eprint = {astro-ph/0603693},
 primaryClass = {astro-ph},
       adsurl = {https://ui.adsabs.harvard.edu/abs/2006ApJ...646...61G}
}

@ARTICLE{Krawczynski2004,
       author = {{Krawczynski}, H. and {Hughes}, S.~B. and {Horan}, D. and {Aharonian}, F. and {Aller}, M.~F. and {Aller}, H. and {Boltwood}, P. and {Buckley}, J. and {Coppi}, P. and {Fossati}, G. and {G{\"o}tting}, N. and {Holder}, J. and {Horns}, D. and {Kurtanidze}, O.~M. and {Marscher}, A.~P. and {Nikolashvili}, M. and {Remillard}, R.~A. and {Sadun}, A. and {Schr{\"o}der}, M.},
        title = "{Multiwavelength Observations of Strong Flares from the TeV Blazar 1ES 1959+650}",
      journal = {\apj},
         year = 2004,
        month = jan,
       volume = {601},
       number = {1},
        pages = {151-164},
          doi = {10.1086/380393},
archivePrefix = {arXiv},
       eprint = {astro-ph/0310158},
 primaryClass = {astro-ph},
       adsurl = {https://ui.adsabs.harvard.edu/abs/2004ApJ...601..151K}
}

@ARTICLE{Newathena2025,
       author = {{Cruise}, Mike and {Guainazzi}, Matteo and {Aird}, James and {Carrera}, Francisco J. and {Costantini}, Elisa and {Corrales}, Lia and {Dauser}, Thomas and {Eckert}, Dominique and {Gastaldello}, Fabio and {Matsumoto}, Hironori and {Osten}, Rachel and {Petrucci}, Pierre-Olivier and {Porquet}, Delphine and {Pratt}, Gabriel W. and {Rea}, Nanda and {Reiprich}, Thomas H. and {Simionescu}, Aurora and {Spiga}, Daniele and {Troja}, Eleonora},
        title = "{The NewAthena mission concept in the context of the next decade of X-ray astronomy}",
      journal = {Nature Astronomy},
         year = 2025,
        month = jan,
       volume = {9},
        pages = {36-44},
          doi = {10.1038/s41550-024-02416-3},
archivePrefix = {arXiv},
       eprint = {2501.03100},
 primaryClass = {astro-ph.IM},
       adsurl = {https://ui.adsabs.harvard.edu/abs/2025NatAs...9...36C}
}

@ARTICLE{EHT2025,
       author = {{Event Horizon Telescope Collaboration} and {Akiyama}, Kazunori and {Albentosa-Ru{\'\i}z}, Ezequiel and {Alberdi}, Antxon and {Alef}, Walter and {Algaba}, Juan Carlos and {Anantua}, Richard and {Asada}, Keiichi and {Azulay}, Rebecca and {Bach}, Uwe and {Baczko}, Anne-Kathrin and {Ball}, David and {Balokovi{\'c}}, Mislav and {Bandyopadhyay}, Bidisha and {Barrett}, John and {Baub{\"o}ck}, Michi and {Benson}, Bradford A. and {Bintley}, Dan and {Blackburn}, Lindy and {Blundell}, Raymond and {Bouman}, Katherine L. and {Bower}, Geoffrey C. and {Bremer}, Michael and {Brissenden}, Roger and {Britzen}, Silke and {Broderick}, Avery E. and {Broguiere}, Dominique and {Bronzwaer}, Thomas and {Bustamante}, Sandra and {Carlos}, Douglas F. and {Carlstrom}, John E. and {Chael}, Andrew and {Chan}, Chi-kwan and {Chang}, Dominic O. and {Chavez}, Erandi and {Chatterjee}, Koushik and {Chatterjee}, Shami and {Chen}, Ming-Tang and {Chen}, Yongjun and {Cheng}, Xiaopeng and {Chichura}, Paul and {Cho}, Ilje and {Christian}, Pierre and {Conroy}, Nicholas S. and {Conway}, John E. and {Crawford}, Thomas M. and {Crew}, Geoffrey B. and {Cruz-Osorio}, Alejandro and {Cui}, Yuzhu and {Curd}, Brandon and {Dahale}, Rohan and {Davelaar}, Jordy and {De Laurentis}, Mariafelicia and {Deane}, Roger and {Desvignes}, Gregory and {Dexter}, Jason and {Dhruv}, Vedant and {Dihingia}, Indu K. and {Doeleman}, Sheperd S. and {Dzib}, Sergio A. and {Eatough}, Ralph P. and {Emami}, Razieh and {Falcke}, Heino and {Farah}, Joseph and {Fish}, Vincent L. and {Fomalont}, Edward and {Alyson Ford}, H. and {Foschi}, Marianna and {Fraga-Encinas}, Raquel and {Freeman}, William T. and {Friberg}, Per and {Fromm}, Christian M. and {Fuentes}, Antonio and {Galison}, Peter and {Gammie}, Charles F. and {Garc{\'\i}a}, Roberto and {Gentaz}, Olivier and {Georgiev}, Boris and {Goddi}, Ciriaco and {Gold}, Roman and {G{\'o}mez-Ruiz}, Arturo I. and {G{\'o}mez}, Jos{\'e} L. and {Gu}, Minfeng and {Gurwell}, Mark and {Hada}, Kazuhiro and {Haggard}, Daryl and {Hesper}, Ronald and {Heumann}, Dirk and {Ho}, Luis C. and {Ho}, Paul and {Hoak}, Dan and {Honma}, Mareki and {Huang}, Chih-Wei L. and {Huang}, Lei and {Hughes}, David H. and {Ikeda}, Shiro and {Violette Impellizzeri}, C.~M. and {Inoue}, Makoto and {Issaoun}, Sara and {James}, David J. and {Jannuzi}, Buell T. and {Janssen}, Michael and {Jeter}, Britton and {Jiang}, Wu and {Jim{\'e}nez-Rosales}, Alejandra and {Johnson}, Michael D. and {Jorstad}, Svetlana and {Jones}, Adam C. and {Joshi}, Abhishek V. and {Jung}, Taehyun and {Karuppusamy}, Ramesh and {Kawashima}, Tomohisa and {Keating}, Garrett K. and {Kettenis}, Mark and {Kim}, Dong-Jin and {Kim}, Jae-Young and {Kim}, Jongsoo and {Kim}, Junhan and {Kino}, Motoki and {Koay}, Jun Yi and {Kocherlakota}, Prashant and {Kofuji}, Yutaro and {Koch}, Patrick M. and {Koyama}, Shoko and {Kramer}, Carsten and {Kramer}, Joana A. and {Kramer}, Michael and {Krichbaum}, Thomas P. and {Kuo}, Cheng-Yu and {La Bella}, Noemi and {Lee}, Deokhyeong and {Lee}, Sang-Sung and {Levis}, Aviad and {Li}, Zhiyuan and {Lico}, Rocco and {Lindahl}, Greg and {Lindqvist}, Michael and {Lisakov}, Mikhail and {Liu}, Jun and {Liu}, Kuo and {Liuzzo}, Elisabetta and {Lo}, Wen-Ping and {Lobanov}, Andrei P. and {Loinard}, Laurent and {Lonsdale}, Colin J. and {Lowitz}, Amy E. and {Lu}, Ru-Sen and {MacDonald}, Nicholas R. and {Mao}, Jirong and {Marchili}, Nicola and {Markoff}, Sera and {Marrone}, Daniel P. and {Marscher}, Alan P. and {Mart{\'\i}-Vidal}, Iv{\'a}n and {Matsushita}, Satoki and {Matthews}, Lynn D. and {Medeiros}, Lia and {Menten}, Karl M. and {Mizuno}, Izumi and {Mizuno}, Yosuke and {Montgomery}, Joshua and {Moran}, James M. and {Moriyama}, Kotaro and {Moscibrodzka}, Monika and {Mulaudzi}, Wanga and {M{\"u}ller}, Cornelia and {M{\"u}ller}, Hendrik and {Mus}, Alejandro and {Musoke}, Gibwa and {Myserlis}, Ioannis and {Nagai}, Hiroshi and {Nagar}, Neil M. and {Nair}, Dhanya G. and {Nakamura}, Masanori and {Narayanan}, Gopal and {Natarajan}, Iniyan and {Nathanail}, Antonios and {Fuentes}, Santiago Navarro and {Neilsen}, Joey and {Ni}, Chunchong and {Nowak}, Michael A. and {Oh}, Junghwan and {Okino}, Hiroki and {S{\'a}nchez}, H{\'e}ctor Ra{\'u}l Olivares and {Oyama}, Tomoaki and {{\"O}zel}, Feryal and {Palumbo}, Daniel C.~M. and {Paraschos}, Georgios Filippos and {Park}, Jongho and {Parsons}, Harriet and {Patel}, Nimesh and {Pen}, Ue-Li and {Pesce}, Dominic W. and {Pi{\'e}tu}, Vincent and {Plavin}, Alexander and {PopStefanija}, Aleksandar and {Porth}, Oliver and {Prather}, Ben and {Principe}, Giacomo and {Psaltis}, Dimitrios},
        title = "{Horizon-scale variability of M87* from 2017─2021 EHT observations}",
      journal = {\aap},
         year = 2025,
        month = dec,
       volume = {704},
          eid = {A91},
        pages = {A91},
          doi = {10.1051/0004-6361/202555855},
archivePrefix = {arXiv},
       eprint = {2509.24593},
 primaryClass = {astro-ph.HE},
       adsurl = {https://ui.adsabs.harvard.edu/abs/2025A&A...704A..91E}
}

@ARTICLE{Weaver2026,
       author = {{Weaver}, Kimberly A. and {Cann}, Jenna M. and {Pfeifle}, Ryan and {McCarthy}, Miranda and {Vega}, Laura D. and {Gamble}, Ron and {Monsue}, Teresa and {Mullaney}, Kyla and {Singha}, Mainak and {Lambrides}, Erini and {McKaig}, Jeffrey and {Carlton}, Isabella and {Whalen}, Kelly and {Kleiner}, Emma and {Mohan}, Atul and {Karmakar}, Subhajeet and {Hornschemeier-Cardiff}, Ann and {Ortiz}, Herbert and {Ricci}, Claudio and {Valencic}, Lynne and {Coleman}, Brandon and {DeGennaro}, Kaylee and {Pandey}, Ruchi},
        title = "{The Need for Ultra High Resolution X-ray Imaging}",
      journal = {arXiv e-prints},
         year = 2026,
        month = jan,
          eid = {arXiv:2601.20823},
        pages = {arXiv:2601.20823},
          doi = {10.48550/arXiv.2601.20823},
archivePrefix = {arXiv},
       eprint = {2601.20823},
 primaryClass = {astro-ph.HE},
       adsurl = {https://ui.adsabs.harvard.edu/abs/2026arXiv260120823W}
}

@INPROCEEDINGS{Uttley2020,
       author = {{Uttley et al.,}, Phil},
        title = "{An x-ray interferometry concept for the ESA Voyage 2050 programme}",
    booktitle = {Society of Photo-Optical Instrumentation Engineers (SPIE) Conference Series},
         year = 2020,
       series = {Society of Photo-Optical Instrumentation Engineers (SPIE) Conference Series},
       volume = {11444},
        month = dec,
          eid = {114441E},
        pages = {114441E},
          doi = {10.1117/12.2562523},
       adsurl = {https://ui.adsabs.harvard.edu/abs/2020SPIE11444E..1EU}
}

@ARTICLE{Gandhi2023,
       author = {{Gandhi}, Poshak},
        title = "{X-ray Astronomy from the Lunar Surface}",
      journal = {arXiv e-prints},
         year = 2023,
        month = oct,
          eid = {arXiv:2310.15215},
        pages = {arXiv:2310.15215},
          doi = {10.48550/arXiv.2310.15215},
archivePrefix = {arXiv},
       eprint = {2310.15215},
 primaryClass = {astro-ph.IM},
       adsurl = {https://ui.adsabs.harvard.edu/abs/2023arXiv231015215G}
}

@InProceedings{AC_review_2018,
author="Chatterjee, Arka",
editor="Mukhopadhyay, Banibrata
and Sasmal, Sudipta",
title="Evolution of Imaging of Black Hole Accretion-Outflow System over Half a Century",
booktitle="Exploring the Universe: From Near Space to Extra-Galactic",
year="2018",
publisher="Springer International Publishing",
address="Cham",
pages="29--38",
isbn="978-3-319-94607-8"
}

@ARTICLE{AC2018,
       author = {{Chatterjee}, Arka and {Chakrabarti}, Sandip K. and {Ghosh}, Himadri and {Garain}, Sudip K.},
        title = "{Images and spectra of time-dependent two-component advective flow in presence of outflows}",
      journal = {\mnras},
         year = 2018,
        month = aug,
       volume = {478},
       number = {3},
        pages = {3356-3366},
          doi = {10.1093/mnras/sty1054},
archivePrefix = {arXiv},
       eprint = {1802.10296},
 primaryClass = {astro-ph.HE},
       adsurl = {https://ui.adsabs.harvard.edu/abs/2018MNRAS.478.3356C}
}

@ARTICLE{AC2017,
       author = {{Chatterjee}, Arka and {Chakrabarti}, Sandip K. and {Ghosh}, Himadri},
        title = "{Images and spectral properties of two-component advective flows around black holes: effects of photon bending}",
      journal = {\mnras},
         year = 2017,
        month = mar,
       volume = {465},
       number = {4},
        pages = {3902-3912},
          doi = {10.1093/mnras/stw2975},
archivePrefix = {arXiv},
       eprint = {1611.06046},
 primaryClass = {astro-ph.HE},
       adsurl = {https://ui.adsabs.harvard.edu/abs/2017MNRAS.465.3902C}
}

@INPROCEEDINGS{GC2004,
       author = {{Gendreau}, Keith C. and {Cash}, Webster C. and {Shipley}, Ann F. and {White}, Nicholas E.},
        title = "{MAXIM x-ray interferometry mission}",
    booktitle = {Optics for EUV, X-Ray, and Gamma-Ray Astronomy},
         year = 2004,
       editor = {{Citterio}, Oberto and {O'Dell}, Stephen L.},
       series = {Society of Photo-Optical Instrumentation Engineers (SPIE) Conference Series},
       volume = {5168},
        month = feb,
        pages = {420-434},
          doi = {10.1117/12.506198},
       adsurl = {https://ui.adsabs.harvard.edu/abs/2004SPIE.5168..420G}
}

@ARTICLE{Cash2003,
       author = {{Cash}, Webster},
        title = "{X-Ray Interferometry}",
      journal = {Experimental Astronomy},
         year = 2003,
        month = mar,
       volume = {16},
       number = {2},
        pages = {91-136},
          doi = {10.1007/s10686-004-2523-5},
       adsurl = {https://ui.adsabs.harvard.edu/abs/2003ExA....16...91C}
}

@ARTICLE{Cash2000,
       author = {{Cash}, Webster and {Shipley}, Ann and {Osterman}, Steve and {Joy}, Marshall},
        title = "{Laboratory detection of X-ray fringes with a grazing-incidence interferometer}",
      journal = {\nat},
         year = 2000,
        month = sep,
       volume = {407},
       number = {6801},
        pages = {160-162},
          doi = {10.1038/35025009},
       adsurl = {https://ui.adsabs.harvard.edu/abs/2000Natur.407..160C}
}

@ARTICLE{Kawaguchi2003,
       author = {{Kawaguchi}, Toshihiro},
        title = "{Comptonization in Super-Eddington Accretion Flow and Growth Timescale of Supermassive Black Holes}",
      journal = {\apj},
         year = 2003,
        month = aug,
       volume = {593},
       number = {1},
        pages = {69-84},
          doi = {10.1086/376404},
archivePrefix = {arXiv},
       eprint = {astro-ph/0304373},
 primaryClass = {astro-ph},
       adsurl = {https://ui.adsabs.harvard.edu/abs/2003ApJ...593...69K}
}

@ARTICLE{JD2024,
       author = {{Jiang}, Yan-Fei and {Dai}, Lixin},
        title = "{Numerical Simulations of Super-Eddington Accretion Flows}",
      journal = {arXiv e-prints},
         year = 2024,
        month = aug,
          eid = {arXiv:2408.16856},
        pages = {arXiv:2408.16856},
          doi = {10.48550/arXiv.2408.16856},
archivePrefix = {arXiv},
       eprint = {2408.16856},
 primaryClass = {astro-ph.HE},
       adsurl = {https://ui.adsabs.harvard.edu/abs/2024arXiv240816856J}
}

@ARTICLE{NIA2003,
       author = {{Narayan}, Ramesh and {Igumenshchev}, Igor V. and {Abramowicz}, Marek A.},
        title = "{Magnetically Arrested Disk: an Energetically Efficient Accretion Flow}",
      journal = {\pasj},
         year = 2003,
        month = dec,
       volume = {55},
        pages = {L69-L72},
          doi = {10.1093/pasj/55.6.L69},
archivePrefix = {arXiv},
       eprint = {astro-ph/0305029},
 primaryClass = {astro-ph},
       adsurl = {https://ui.adsabs.harvard.edu/abs/2003PASJ...55L..69N}
}

@ARTICLE{JSD2019,
       author = {{Jiang}, Yan-Fei and {Stone}, James M. and {Davis}, Shane W.},
        title = "{Super-Eddington Accretion Disks around Supermassive Black Holes}",
      journal = {\apj},
         year = 2019,
        month = aug,
       volume = {880},
       number = {2},
          eid = {67},
        pages = {67},
          doi = {10.3847/1538-4357/ab29ff},
archivePrefix = {arXiv},
       eprint = {1709.02845},
 primaryClass = {astro-ph.HE},
       adsurl = {https://ui.adsabs.harvard.edu/abs/2019ApJ...880...67J}
}

@ARTICLE{Madau1988,
       author = {{Madau}, Piero},
        title = "{Thick Accretion Disks around Black Holes and the UV/Soft X-Ray Excess in Quasars}",
      journal = {\apj},
         year = 1988,
        month = apr,
       volume = {327},
        pages = {116},
          doi = {10.1086/166175},
       adsurl = {https://ui.adsabs.harvard.edu/abs/1988ApJ...327..116M}
}

@ARTICLE{Jaroszynski1986,
       author = {{Jaroszynski}, M.},
        title = "{Oscillations of thick accretion discs}",
      journal = {\mnras},
         year = 1986,
        month = jun,
       volume = {220},
        pages = {869-881},
          doi = {10.1093/mnras/220.4.869},
       adsurl = {https://ui.adsabs.harvard.edu/abs/1986MNRAS.220..869J}
}

@ARTICLE{SKC1985b,
       author = {{Chakrabarti}, S.~K.},
        title = "{Analytic structure of cosmic radio jets - A preliminary investigation}",
      journal = {\apj},
         year = 1985,
        month = jan,
       volume = {288},
        pages = {7-13},
          doi = {10.1086/162756},
       adsurl = {https://ui.adsabs.harvard.edu/abs/1985ApJ...288....7C}
}

@ARTICLE{SKC1985,
       author = {{Chakrabarti}, S.~K.},
        title = "{The natural angular momentum distribution in the study of thick disks around black holes}",
      journal = {\apj},
         year = 1985,
        month = jan,
       volume = {288},
        pages = {1-6},
          doi = {10.1086/162755},
       adsurl = {https://ui.adsabs.harvard.edu/abs/1985ApJ...288....1C}
}

@ARTICLE{Fukue1982,
       author = {{Fukue}, Jun},
        title = "{Jets from a Geometrically Thick Disk}",
      journal = {\pasj},
         year = 1982,
        month = jul,
       volume = {34},
       number = {2},
        pages = {163-171},
          doi = {10.1093/pasj/34.2.163},
       adsurl = {https://ui.adsabs.harvard.edu/abs/1982PASJ...34..163F}
}

@ARTICLE{RBBP1982,
       author = {{Rees}, M.~J. and {Begelman}, M.~C. and {Blandford}, R.~D. and {Phinney}, E.~S.},
        title = "{Ion-supported tori and the origin of radio jets}",
      journal = {\nat},
         year = 1982,
        month = jan,
       volume = {295},
       number = {5844},
        pages = {17-21},
          doi = {10.1038/295017a0},
       adsurl = {https://ui.adsabs.harvard.edu/abs/1982Natur.295...17R}
}

@ARTICLE{Sikora1981,
       author = {{Sikora}, M.},
        title = "{Superluminous Accretion Discs}",
      journal = {\mnras},
         year = 1981,
        month = jul,
       volume = {196},
        pages = {257},
          doi = {10.1093/mnras/196.2.257},
       adsurl = {https://ui.adsabs.harvard.edu/abs/1981MNRAS.196..257S}
}

@ARTICLE{ACN1980,
       author = {{Abramowicz}, M.~A. and {Calvani}, M. and {Nobili}, L.},
        title = "{Thick accretion disks with super-Eddington luminosities}",
      journal = {\apj},
         year = 1980,
        month = dec,
       volume = {242},
        pages = {772-788},
          doi = {10.1086/158512},
       adsurl = {https://ui.adsabs.harvard.edu/abs/1980ApJ...242..772A}
}

@ARTICLE{Ulmer1997,
       author = {{Ulmer}, Andrew},
        title = "{Evolution of Thick Accretion Disks Produced by Tidal Disruption Events}",
      journal = {arXiv e-prints},
         year = 1997,
        month = aug,
          eid = {astro-ph/9708265},
        pages = {astro-ph/9708265},
          doi = {10.48550/arXiv.astro-ph/9708265},
archivePrefix = {arXiv},
       eprint = {astro-ph/9708265},
 primaryClass = {astro-ph},
       adsurl = {https://ui.adsabs.harvard.edu/abs/1997astro.ph..8265U}
}

@ARTICLE{PW1980,
       author = {{Paczy{\'n}sky}, B. and {Wiita}, P.~J.},
        title = "{Thick Accretion Disks and Supercritical Luminosities}",
      journal = {\aap},
         year = 1980,
        month = aug,
       volume = {88},
        pages = {23},
       adsurl = {https://ui.adsabs.harvard.edu/abs/1980A&A....88...23P}
}

@ARTICLE{Kozlowski1978,
       author = {{Kozlowski}, M. and {Jaroszynski}, M. and {Abramowicz}, M.~A.},
        title = "{The analytic theory of fluid disks orbiting the Kerr black hole.}",
      journal = {\aap},
         year = 1978,
        month = feb,
       volume = {63},
       number = {1-2},
        pages = {209-220},
       adsurl = {https://ui.adsabs.harvard.edu/abs/1978A&A....63..209K}
}

@ARTICLE{Abramowicz1978,
       author = {{Abramowicz}, M. and {Jaroszynski}, M. and {Sikora}, M.},
        title = "{Relativistic, accreting disks.}",
      journal = {\aap},
         year = 1978,
        month = feb,
       volume = {63},
        pages = {221-224},
       adsurl = {https://ui.adsabs.harvard.edu/abs/1978A&A....63..221A}
}

@article{FM2013,
  author = {{Foreman-Mackey}, D. and {Hogg}, D.~W. and {Lang}, D. and {Goodman}, J.},
  title = "{emcee: The MCMC Hammer}",
  journal = {\pasp},
  year = 2013,
  volume = 125,
  number = 925,
  pages = {306-312},
  doi = {10.1086/670067},
  adsurl = {https://ui.adsabs.harvard.edu/abs/2013PASP..125..306F}
}

@misc{Chakraborty2026,
      title={X-ray Spectral-Timing Properties of Tidal Disruption Events}, 
      author={Joheen Chakraborty and Megan Masterson and Andrew Mummery and Erin Kara and Christos Panagiotou and Riccardo Arcodia and Vera Berger},
      year={2026},
      eprint={2602.16868},
      archivePrefix={arXiv},
      primaryClass={astro-ph.HE},
      url={https://arxiv.org/abs/2602.16868}, 
}

@ARTICLE{Koss2025,
       author = {{Koss}, Michael and {Aftab}, Nafisa and {Allen}, Steven W. and {Amato}, Roberta and {An}, Hongjun and {Andreoni}, Igor and {Anguita}, Timo and {Arcodia}, Riccardo and {Ayres}, Thomas and {Bachetti}, Matteo and {Baglio}, Maria Cristina and {Bahramian}, Arash and {Balboni}, Marco and {Baldi}, Ranieri D. and {Balman}, Solen and {Bamba}, Aya and {Banados}, Eduardo and {Bao}, Tong and {Bartalucci}, Iacopo and {Basu-Zych}, Antara and {Batalha}, Rebeca and {Battistini}, Lorenzo and {Bauer}, Franz Erik and {Beardmore}, Andy and {Becker}, Werner and {Behar}, Ehud and {Belfiore}, Andrea and {Beniamini}, Paz and {Bertola}, Elena and {Bessa}, Vinicius and {Best}, Henry and {Bianchi}, Stefano and {Biava}, N. and {Binder}, Breanna A. and {Blanton}, Elizabeth L. and {Bodaghee}, Arash and {Bogdanovic}, Tamara and {Bogensberger}, David and {Bonafede}, A. and {Bonetti}, Matteo and {Bordas}, Pol and {Borghese}, Alice and {Botteon}, Andrea and {Boula}, Stella and {Bozzo}, Enrico and {Branchesi}, Marica and {Brandt}, William Nielsen and {Bregman}, Joel and {Brighenti}, Fabrizio and {Bronzini}, Ettore and {Brunelli}, Giulia and {Brusa}, Marcella and {Bulbul}, Esra and {Burdge}, Kevin and {Caccianiga}, Alessandro and {Calzadilla}, Michael and {Campana}, Sergio and {Capalbi}, Milvia and {Capitanio}, Fiamma and {Cappelluti}, Nico and {Carney}, Jonathan and {Casanova}, Sabrina and {Castro}, Daniel and {Cenko}, S. Bradley and {Chakraborty}, Joheen and {Chakraborty}, Priyanka and {Chartas}, George and {Chatterjee}, Arka and {Choudhury}, Prakriti Pal and {Cilley}, Raven and {Civano}, Francesca and {Comastri}, Andrea and {Connor}, Thomas and {Corcoran}, Michael F. and {Corrales}, Lia and {Coti Zelati}, Francesco and {Cui}, Weiguang and {D'Ammando}, Filippo and {Dage}, Kristen and {Daylan}, Tansu and {De Grandi}, Sabrina and {De Rosa}, Alessandra and {Decarli}, Roberto and {Decourchelle}, Anne and {Degenaar}, Nathalie and {Del Popolo}, Antonino and {Di Marco}, Alessandro and {Di Salvo}, Tiziana and {Dichiara}, Simone and {DiKerby}, Stephen and {Dillmann}, Steven and {Doerksen}, Neil and {Draghis}, Paul and {Drake}, Jeremy J. and {Ducci}, Lorenzo and {Dupke}, Renato and {Durbak}, Joseph and {Duvvuri}, Girish M. and {Dykaar}, Hannah and {Eckert}, Dominique and {Elvis}, Martin and {Espaillat}, Catherine and {Esposito}, Paolo and {Furst}, Felix and {Fabbiano}, Giuseppina and {Fagin}, Joshua and {Falcone}, Abraham and {Fedorova}, Elena and {Feinstein}, Adina and {Fernandez Fernandez}, Jorge and {Ferrand}, Gilles and {Flores}, Anthony M. and {Foo}, N. and {Foo}, Nicholas and {Foord}, Adi and {Franchini}, Alessia and {Fraschetti}, Federico and {Frye}, Brenda L. and {Lowenthal}, James D. and {Fryer}, Chris and {Fujimoto}, Shin-ichiro and {Gagnon}, Seth and {Gallo}, Luigi and {Garcia Diaz}, Carlos and {Gaspari}, Massimo and {Gastaldello}, Fabio and {Gelfand}, Joseph D. and {Gezari}, Suvi and {Ghizzardi}, Simona and {Giacintucci}, Simona and {Gill}, A. and {Gilli}, Roberto and {Gitti}, Myriam and {Giustini}, Margherita and {Gnarini}, Andrea and {Grandi}, Paola and {Gross}, Arran and {Gu}, Liyi and {Gunderson}, Sean and {Gunther}, Hans Moritz and {Haggard}, Daryl and {Hamaguchi}, Kenji and {Hare}, Jeremy and {Harrington}, Kevin C. and {Heinke}, Craig and {Heinz}, Sebastian and {Hlavacek-Larrondo}, Julie and {Ho}, Wynn C.~G. and {Hodges-Kluck}, Edmund and {Homan}, Jeroen and {Huang}, R. and {Ighina}, Luca and {Ignesti}, Alessandro and {Imbrogno}, Matteo and {Irwin}, Christopher and {Irwin}, Jimmy and {Islam}, Nazma and {Israel}, Gian Luca and {Jacobson-Galan}, Wynn and {Jain}, Chetana and {Jana}, Arghajit and {Jaodand}, Amruta and {Jennings}, Fred and {Jiang}, Jiachen and {Jimenez-Andrade}, Eric F. and {Jimenez-Teja}, Y. and {Johnson}, S. and {Jonker}, Peter and {Kamieneski}, Patrick S. and {Kammoun}, Elias and {Kara}, Erin and {Kargaltsev}, Oleg and {King}, George W. and {Kirmizibayrak}, Demet and {Klingler}, Noel and {Kong}, Albert K.~H. and {Kounkel}, Marina and {Kumar}, Manish and {Kutyrev}, Alexander and {Kyer}, Rebecca and {La Monaca}, Fabio and {Lambrides}, Erini and {Lanzuisi}, Giorgio and {Lee}, Wonki and {Lehmer}, Bret and {Lentini}, Elisa and {Lepore}, Marika and {Li}, Jiangtao and {Lisse}, Carey M. and {Liu}, Daizhong and {Liu}, Tingting and {Isla Llave}, Monica and {Locatelli}, Nicola and {Lopez}, Laura A. and {Lopez}, Sebastian and {Lovisari}, Lorenzo and {Lusso}, Elisabeta and {Mac Intyre}, Brydyn and {MacMaster}, Austin and {Maiolino}, Roberto},
        title = "{The Advanced X-ray Imaging Satellite (AXIS) Community Science Book}",
      journal = {arXiv e-prints},
         year = 2025,
        month = oct,
          eid = {arXiv:2511.00253},
        pages = {arXiv:2511.00253},
          doi = {10.48550/arXiv.2511.00253},
archivePrefix = {arXiv},
       eprint = {2511.00253},
 primaryClass = {astro-ph.HE},
       adsurl = {https://ui.adsabs.harvard.edu/abs/2025arXiv251100253K}
}

@ARTICLE{Peterson2004,
       author = {{Peterson}, B.~M. and {Ferrarese}, L. and {Gilbert}, K.~M. and {Kaspi}, S. and {Malkan}, M.~A. and {Maoz}, D. and {Merritt}, D. and {Netzer}, H. and {Onken}, C.~A. and {Pogge}, R.~W. and {Vestergaard}, M. and {Wandel}, A.},
        title = "{Central Masses and Broad-Line Region Sizes of Active Galactic Nuclei. II. A Homogeneous Analysis of a Large Reverberation-Mapping Database}",
      journal = {\apj},
         year = 2004,
        month = oct,
       volume = {613},
       number = {2},
        pages = {682-699},
          doi = {10.1086/423269},
archivePrefix = {arXiv},
       eprint = {astro-ph/0407299},
 primaryClass = {astro-ph},
       adsurl = {https://ui.adsabs.harvard.edu/abs/2004ApJ...613..682P}
}

@ARTICLE{2002ApJ...572L.173L,
       author = {{Liu}, B.~F. and {Mineshige}, S. and {Shibata}, K.},
        title = "{A Simple Model for a Magnetic Reconnection-heated Corona}",
      journal = {\apjl},
         year = 2002,
        month = jun,
       volume = {572},
       number = {2},
        pages = {L173-L176},
          doi = {10.1086/341877},
archivePrefix = {arXiv},
       eprint = {astro-ph/0205257},
 primaryClass = {astro-ph},
       adsurl = {https://ui.adsabs.harvard.edu/abs/2002ApJ...572L.173L}
}

@article{cendes_radio_2021,
	title = {Radio {Observations} of an {Ordinary} {Outflow} from the {Tidal} {Disruption} {Event} {AT2019dsg}},
	volume = {919},
	issn = {0004-637X},
	url = {https://ui.adsabs.harvard.edu/abs/2021ApJ...919..127C},
	doi = {10.3847/1538-4357/ac110a},
	urldate = {2022-05-07},
	journal = {The Astrophysical Journal},
	author = {Cendes, Y. and Alexander, K. D. and Berger, E. and Eftekhari, T. and Williams, P. K. G. and Chornock, R.},
	month = oct,
	year = {2021},
	note = {ADS Bibcode: 2021ApJ...919..127C},
	pages = {127},
}

@article{alexander_radio_2020,
	title = {Radio {Properties} of {Tidal} {Disruption} {Events}},
	volume = {216},
	issn = {0038-6308, 1572-9672},
	url = {https://link.springer.com/10.1007/s11214-020-00702-w},
	doi = {10.1007/s11214-020-00702-w},
	language = {en},
	number = {5},
	urldate = {2022-04-29},
	journal = {Space Sci Rev},
	author = {Alexander, Kate D. and van Velzen, Sjoert and Horesh, Assaf and Zauderer, B. Ashley},
	month = aug,
	year = {2020},
	pages = {81},
}

@ARTICLE{2020NewAR..8901538D,
       author = {{De Colle}, Fabio and {Lu}, Wenbin},
        title = "{Jets from Tidal Disruption Events}",
      journal = {\nar},
         year = 2020,
        month = sep,
       volume = {89},
          eid = {101538},
        pages = {101538},
          doi = {10.1016/j.newar.2020.101538},
archivePrefix = {arXiv},
       eprint = {1911.01442},
 primaryClass = {astro-ph.HE},
       adsurl = {https://ui.adsabs.harvard.edu/abs/2020NewAR..8901538D}
}

@ARTICLE{James2013,
       author = {{Guillochon}, James and {Ramirez-Ruiz}, Enrico},
        title = "{Hydrodynamical Simulations to Determine the Feeding Rate of Black Holes by the Tidal Disruption of Stars: The Importance of the Impact Parameter and Stellar Structure}",
      journal = {\apj},
         year = 2013,
        month = apr,
       volume = {767},
       number = {1},
          eid = {25},
        pages = {25},
          doi = {10.1088/0004-637X/767/1/25},
archivePrefix = {arXiv},
       eprint = {1206.2350},
 primaryClass = {astro-ph.HE},
       adsurl = {https://ui.adsabs.harvard.edu/abs/2013ApJ...767...25G}
}

@ARTICLE{Rossi+2021,
       author = {{Rossi}, E.~M. and {Stone}, N.~C. and {Law-Smith}, J.~A.~P. and {Macleod}, M. and {Lodato}, G. and {Dai}, J.~L. and {Mandel}, I.},
        title = "{The Process of Stellar Tidal Disruption by Supermassive Black Holes}",
      journal = {\ssr},
         year = 2021,
        month = apr,
       volume = {217},
       number = {3},
          eid = {40},
        pages = {40},
          doi = {10.1007/s11214-021-00818-7},
archivePrefix = {arXiv},
       eprint = {2005.12528},
 primaryClass = {astro-ph.HE},
       adsurl = {https://ui.adsabs.harvard.edu/abs/2021SSRv..217...40R}
}

@ARTICLE{PGKH2020,
       author = {{Park}, Gwanwoo and {Hayasaki}, Kimitake},
        title = "{Tidal Disruption Flares from Stars on Marginally Bound and Unbound Orbits}",
      journal = {\apj},
         year = 2020,
        month = sep,
       volume = {900},
       number = {1},
          eid = {3},
        pages = {3},
          doi = {10.3847/1538-4357/ab9ebb},
archivePrefix = {arXiv},
       eprint = {2001.04548},
 primaryClass = {astro-ph.HE},
       adsurl = {https://ui.adsabs.harvard.edu/abs/2020ApJ...900....3P}
}

@ARTICLE{2023ApJ...959...19Z,
       author = {{Zhong}, Shiyan and {Hayasaki}, Kimitake and {Li}, Shuo and {Berczik}, Peter and {Spurzem}, Rainer},
        title = "{Exploring the Origin of Stars on Bound and Unbound Orbits Causing Tidal Disruption Events}",
      journal = {\apj},
         year = 2023,
        month = dec,
       volume = {959},
       number = {1},
          eid = {19},
        pages = {19},
          doi = {10.3847/1538-4357/ad0122},
archivePrefix = {arXiv},
       eprint = {2011.09400},
 primaryClass = {astro-ph.HE},
       adsurl = {https://ui.adsabs.harvard.edu/abs/2023ApJ...959...19Z}
}

@ARTICLE{Sun2015,
       author = {{Sun}, Hui and {Zhang}, Bing and {Li}, Zhuo},
        title = "{Extragalactic High-energy Transients: Event Rate Densities and Luminosity Functions}",
      journal = {\apj},
         year = 2015,
        month = oct,
       volume = {812},
       number = {1},
          eid = {33},
        pages = {33},
          doi = {10.1088/0004-637X/812/1/33},
archivePrefix = {arXiv},
       eprint = {1509.01592},
 primaryClass = {astro-ph.HE},
       adsurl = {https://ui.adsabs.harvard.edu/abs/2015ApJ...812...33S}
}

@ARTICLE{KH2013,
       author = {{Kormendy}, John and {Ho}, Luis C.},
        title = "{Coevolution (Or Not) of Supermassive Black Holes and Host Galaxies}",
      journal = {\araa},
         year = 2013,
        month = aug,
       volume = {51},
       number = {1},
        pages = {511-653},
          doi = {10.1146/annurev-astro-082708-101811},
archivePrefix = {arXiv},
       eprint = {1304.7762},
 primaryClass = {astro-ph.CO},
       adsurl = {https://ui.adsabs.harvard.edu/abs/2013ARA&A..51..511K}
}

@ARTICLE{Kara2016,
       author = {{Kara}, Erin and {Miller}, Jon M. and {Reynolds}, Chris and {Dai}, Lixin},
        title = "{Relativistic reverberation in the accretion flow of a tidal disruption event}",
      journal = {\nat},
         year = 2016,
        month = jul,
       volume = {535},
       number = {7612},
        pages = {388-390},
          doi = {10.1038/nature18007},
archivePrefix = {arXiv},
       eprint = {1606.06736},
 primaryClass = {astro-ph.HE},
       adsurl = {https://ui.adsabs.harvard.edu/abs/2016Natur.535..388K}
}

@ARTICLE{Andreoni2022,
       author = {{Andreoni}, Igor and {Coughlin}, Michael W. and {Perley}, Daniel A. and {Yao}, Yuhan and {Lu}, Wenbin and {Cenko}, S. Bradley and {Kumar}, Harsh and {Anand}, Shreya and {Ho}, Anna Y.~Q. and {Kasliwal}, Mansi M. and {de Ugarte Postigo}, Antonio and {Sagu{\'e}s-Carracedo}, Ana and {Schulze}, Steve and {Kann}, D. Alexander and {Kulkarni}, S.~R. and {Sollerman}, Jesper and {Tanvir}, Nial and {Rest}, Armin and {Izzo}, Luca and {Somalwar}, Jean J. and {Kaplan}, David L. and {Ahumada}, Tom{\'a}s and {Anupama}, G.~C. and {Auchettl}, Katie and {Barway}, Sudhanshu and {Bellm}, Eric C. and {Bhalerao}, Varun and {Bloom}, Joshua S. and {Bremer}, Michael and {Bulla}, Mattia and {Burns}, Eric and {Campana}, Sergio and {Chandra}, Poonam and {Charalampopoulos}, Panos and {Cooke}, Jeff and {D'Elia}, Valerio and {Das}, Kaustav Kashyap and {Dobie}, Dougal and {Ag{\"u}{\'\i} Fern{\'a}ndez}, Jos{\'e} Feliciano and {Freeburn}, James and {Fremling}, Cristoffer and {Gezari}, Suvi and {Goode}, Simon and {Graham}, Matthew J. and {Hammerstein}, Erica and {Karambelkar}, Viraj R. and {Kilpatrick}, Charles D. and {Kool}, Erik C. and {Krips}, Melanie and {Laher}, Russ R. and {Leloudas}, Giorgos and {Levan}, Andrew and {Lundquist}, Michael J. and {Mahabal}, Ashish A. and {Medford}, Michael S. and {Miller}, M. Coleman and {M{\"o}ller}, Anais and {Mooley}, Kunal P. and {Nayana}, A.~J. and {Nir}, Guy and {Pang}, Peter T.~H. and {Paraskeva}, Emmy and {Perley}, Richard A. and {Petitpas}, Glen and {Pursiainen}, Miika and {Ravi}, Vikram and {Ridden-Harper}, Ryan and {Riddle}, Reed and {Rigault}, Mickael and {Rodriguez}, Antonio C. and {Rusholme}, Ben and {Sharma}, Yashvi and {Smith}, I.~A. and {Stein}, Robert D. and {Th{\"o}ne}, Christina and {Tohuvavohu}, Aaron and {Valdes}, Frank and {van Roestel}, Jan and {Vergani}, Susanna D. and {Wang}, Qinan and {Zhang}, Jielai},
        title = "{A very luminous jet from the disruption of a star by a massive black hole}",
      journal = {\nat},
         year = 2022,
        month = dec,
       volume = {612},
       number = {7940},
        pages = {430-434},
          doi = {10.1038/s41586-022-05465-8},
archivePrefix = {arXiv},
       eprint = {2211.16530},
 primaryClass = {astro-ph.HE},
       adsurl = {https://ui.adsabs.harvard.edu/abs/2022Natur.612..430A}
}

@misc{Caiazzo2019,
  author       = {Caiazzo, Ilaria and
                  Belloni, Tomaso and
                  Cackett, Edward and
                  Damascelli, Andrea and
                  De Rosa, Alessandra and
                  Dosanjh, Pinder and
                  Feroci, Marco and
                  Gallagher, Sarah and
                  Gallo, Luigi and
                  Guest, Benson and
                  Haggard, Daryl and
                  Heinke, Craig and
                  Heyl, Jeremy and
                  Hoffman, Kelsey and
                  Ingram, Adam R. and
                  Kırmızıbayrak, Demet and
                  Marshall, Herman and
                  Morsink, Sharon and
                  Rau, Wolfgang and
                  Ripoche, Paul and
                  Safi-Harb, Samar and
                  Sivakoff, Gregory R. and
                  Stairs, Ingrid and
                  Stella, Luigi and
                  Swetz, Daniel S. and
                  Ullom, Joel N.},
  title        = {{Unveiling the secrets of black holes and neutron 
                   stars with high-throughput, high-energy resolution
                   X-ray spectroscopy}},
  month        = oct,
  year         = 2019,
  note         = {White paper identifier W036},
  publisher    = {Zenodo},
  doi          = {10.5281/zenodo.3824441},
  url          = {https://doi.org/10.5281/zenodo.3824441}
}

@INPROCEEDINGS{Heyl2019,
       author = {{Heyl}, Jeremy and {Caiazzo}, Ilaria and {Hoffman}, Kelsey and {Gallagher}, Sarah and {Safi-Harb}, Samar and {Damascelli}, Andrea and {Dosanjh}, Pinder and {Gallo}, Luigi and {Haggard}, Daryl and {Heinke}, Craig and {Kirmizibayrak}, Demet and {Morsink}, Sharon and {Rau}, Wolfgang and {Ripoche}, Paul and {Sivakoff}, Gregory R. and {Stairs}, Ingrid and {Belloni}, Tomaso and {Cackett}, Edward and {De Rosa}, Alessandra and {Feroci}, Marco and {Ingram}, Adam R. and {Marshall}, Herman and {Stella}, Luigi and {Swetz}, Daniel S. and {Ullom}, Joel N.},
        title = "{The Colibr{\`\i} High-Resolution X-ray Telescope}",
    booktitle = {Bulletin of the American Astronomical Society},
         year = 2019,
       volume = {51},
        month = sep,
          eid = {175},
        pages = {175},
       adsurl = {https://ui.adsabs.harvard.edu/abs/2019BAAS...51g.175H}
}

@inproceedings{Tashiro18,
author = {Makoto Tashiro and Hironori Maejima and Kenichi Toda and Richard Kelley and Lillian Reichenthal and James Lobell and Robert Petre and Matteo Guainazzi and Elisa Costantini and Mark Edison and Ryuichi Fujimoto and Martin  Grim and Kiyoshi Hayashida and Jan-Willem den Herder and Yoshitaka Ishisaki and Stéphane Paltani and Kyoko Matsushita and Koji Mori and Gary Sneiderman and Yoh Takei and Yukikatsu Terada and Hiroshi Tomida and Hiroki Akamatsu and Lorella Angelini and Yoshitaka Arai and Hisamitsu Awaki and Lurli Babyk and Aya Bamba and Peter Barfknecht and Kim Barnstable and Thomas Bialas and Branimir Blagojevic and Joseph Bonafede and Clifford Brambora and Laura Brenneman and Greg Brown and Kimberly Brown and Laura Burns and Edgar Canavan and Tim Carnahan and Meng Chiao and Brian Comber and Lia Corrales and Cor de Vries and Johannes Dercksen and Maria Diaz-Trigo and Tyrone Dillard and Michael DiPirro and Chris Done and Tadayasu Dotani and Ken Ebisawa and Megan Eckart and Teruaki Enoto and Yuichiro Ezoe and Carlo Ferrigno and Yasushi Fukazawa and Yutaka Fujita and Akihiro Furuzawa and Luigi Gallo and Steve Graham and Liyi Gu and Kohichi Hagino and Kenji Hamaguchi and Isamu Hatsukade and Dean Hawes and Takayuki Hayashi and Cailey Hegarty and Natalie Hell and Junko Hiraga and Edmund Hodges-Kluck and Matt Holland and Ann Hornschemeier and Akio Hoshino and Yuto Ichinohe and Ryo Iizuka and Kazunori Ishibashi and Manabu Ishida and Kumi Ishikawa and Kosei Ishimura and Bryan James and Timothy Kallman and Erin Kara and Satoru Katsuda and Steven Kenyon and Caroline Kilbourne and Mark Kimball and Takao Kitaguti and Shunji Kitamoto and Shogo Kobayashi and Takayoshi Kohmura and Shu Koyama and Aya Kubota and Maurice Leutenegger and Tom Lockard and Mike Loewenstein and Yoshitomo Maeda and Lynette Marbley and Maxim Markevitch and Hironori Matsumoto and Keiichi Matsuzaki and Dan McCammon and Brian McNamara and Joseph Miko and Eric Miller and Jon Miller and Kenji Minesugi and Ikuyuki Mitsuishi and Tsunefumi Mizuno and Hideyuki Mori and Koji Mukai and Hiroshi Murakami and Richard Mushotzky and Hiroshi Nakajima and Hideto Nakamura and Shinya Nakashima and Kazuhiro Nakazawa and Chikara Natsukari and Kenichiro Nigo and Yusuke Nishioka and Kumiko Nobukawa and Masayoshi Nobukawa and Hirofumi Noda and Hirokazu Odaka and Mina Ogawa and Takaya Ohashi and Masahiro Ohno and Masayuki Ohta and Takashi Okajima and Atsushi Okamoto and Michitaka Onizuka and Naomi Ota and Masanobu Ozaki and Paul Plucinsky and F. Scott Porter and Katja Pottschmidt and Kosuke Sato and Rie Sato and Makoto Sawada and Hiromi Seta and Ken Shelton and Yasuko Shibano and Maki Shida and Megumi Shidatsu and Peter Shirron and Aurora Simionescu and Randall Smith and Kazunori Someya and Yang Soong and Yasuharu Suagawara and Andy Szymkowiak and Hiromitsu Takahashi and Toru Tamagawa and Takayuki Tamura and Takaaki Tanaka and Yuichi Terashima and Yohko Tsuboi and Masahiro Tsujimoto and Hiroshi Tsunemi and Takeshi Tsuru and Hiroyuki Uchida and Hideki Uchiyama and Yoshihiro Ueda and Shinichiro Uno and Thomas Walsh and Shin Watanabe and Brian Williams and Rob Wolfs and Michael Wright and Shinya Yamada and Hiroya Yamaguchi and Kazutaka Yamaoka and Noriko Yamasaki and Shigeo Yamauchi and Makoto Yamauchi and Keiichi Yanagase and Tahir Yaqoob and Susumu Yasuda and Nasa Yoshioka and Jaime Zabala and Zhuravleva Irina},
title = {{Concept of the X-ray Astronomy Recovery Mission}},
volume = {10699},
booktitle = {Space Telescopes and Instrumentation 2018: Ultraviolet to Gamma Ray},
editor = {Jan-Willem A. den Herder and Shouleh Nikzad and Kazuhiro Nakazawa},
organization = {International Society for Optics and Photonics},
publisher = {SPIE},
pages = {520 -- 531},
year = {2018},
doi = {10.1117/12.2309455},
URL = {https://doi.org/10.1117/12.2309455}
}

@ARTICLE{SSH2023,
       author = {{Safi-Harb}, S. and {Burdge}, K.~B. and {Bodaghee}, A. and {An}, H. and {Guest}, B. and {Hare}, J. and {Hebbar}, P. and {Ho}, W.~C.~G. and {Kargaltsev}, O. and {Kirmizibayrak}, D. and {Klingler}, N. and {Nynka}, M. and {Reynolds}, M.~T. and {Sasaki}, M. and {Sridhar}, N. and {Vasilopoulos}, G. and {Woods}, T.~E. and {Yang}, H. and {Heinke}, C. and {Kong}, A. and {Li}, J. and {MacMaster}, A. and {Mallick}, L. and {Treyturik}, C. and {Tsuji}, N. and {Binder}, B. and {Braun}, C. and {Chang}, H. -K. and {Chatterjee}, A. and {Ferrand}, G. and {Holland-Ashford}, T. and {Ng}, C. -Y. and {Plotkin}, R. and {Romani}, R. and {Zhang}, S.},
        title = "{From Stellar Death to Cosmic Revelations: Zooming in on Compact Objects, Relativistic Outflows and Supernova Remnants with AXIS}",
      journal = {arXiv e-prints},
         year = 2023,
        month = nov,
          eid = {arXiv:2311.07673},
        pages = {arXiv:2311.07673},
          doi = {10.48550/arXiv.2311.07673},
archivePrefix = {arXiv},
       eprint = {2311.07673},
 primaryClass = {astro-ph.HE},
       adsurl = {https://ui.adsabs.harvard.edu/abs/2023arXiv231107673S}
}

@ARTICLE{Reynolds2023,
       author = {{Reynolds}, Christopher S. and {Kara}, Erin A. and {Mushotzky}, Richard F. and {Ptak}, Andrew and {Koss}, Michael J. and {Williams}, Brian J. and {Allen}, Steven W. and {Bauer}, Franz E. and {Bautz}, Marshall and {Bodaghee}, Arash and {Burdge}, Kevin B. and {Cappelluti}, Nico and {Cenko}, Brad and {Chartas}, George and {Chan}, Kai-Wing and {Corrales}, L{\'\i}a and {Daylan}, Tansu and {Falcone}, Abraham D. and {Foord}, Adi and {Grant}, Catherine E. and {Habouzit}, M{\'e}lanie and {Haggard}, Daryl and {Herrmann}, Sven and {Hodges-Kluck}, Edmund and {Kargaltsev}, Oleg and {King}, George W. and {Kounkel}, Marina and {Lopez}, Laura A. and {Marchesi}, Stefano and {McDonald}, Michael and {Meyer}, Eileen and {Miller}, Eric D. and {Nynka}, Melania and {Okajima}, Takashi and {Pacucci}, Fabio and {Russell}, Helen R. and {Safi-Harb}, Samar and {Stassun}, Keivan G. and {Trindade Falc{\~a}o}, Anna and {Walker}, Stephen A. and {Wilms}, Joern and {Yukita}, Mihoko and {Zhang}, William W.},
        title = "{Overview of the Advanced X-ray Imaging Satellite (AXIS)}",
      journal = {arXiv e-prints},
         year = 2023,
        month = nov,
          eid = {arXiv:2311.00780},
        pages = {arXiv:2311.00780},
          doi = {10.48550/arXiv.2311.00780},
archivePrefix = {arXiv},
       eprint = {2311.00780},
 primaryClass = {astro-ph.IM},
       adsurl = {https://ui.adsabs.harvard.edu/abs/2023arXiv231100780R}
}

@INPROCEEDINGS{Mushotzky2019,
       author = {{Mushotzky}, R. and {AXIS Team}},
        title = "{AXIS a High Spatial Resolution X-Ray Probe Mission Study}",
    booktitle = {The Space Astrophysics Landscape for the 2020s and Beyond},
         year = 2019,
       editor = {{Moores J.~E.} and {King P.~L.} and {Smith C.~L.} and {Martinez G.~M.} and {Newman C.~E.} and {Guzewich S.~D.} and {Meslin P. -Y.} and {Webster C.~R.} and {Mahaffy P.~R.} and {Atreya S.~K.} and {Schuerger A.~C.}},
       series = {LPI Contributions},
       volume = {2135},
        month = apr,
          eid = {5025},
        pages = {5025},
       adsurl = {https://ui.adsabs.harvard.edu/abs/2019LPICo2135.5025M}
}

@ARTICLE{Patra2019,
       author = {{Patra}, Dusmanta and {Chatterjee}, Arka and {Dutta}, Broja G. and {Chakrabarti}, Sandip K. and {Nandi}, Prantik},
        title = "{Evidence of Outflow-induced Soft Lags of Galactic Black Holes}",
      journal = {\apj},
         year = 2019,
        month = dec,
       volume = {886},
       number = {2},
          eid = {137},
        pages = {137},
          doi = {10.3847/1538-4357/ab4c34},
archivePrefix = {arXiv},
       eprint = {1901.02245},
 primaryClass = {astro-ph.HE},
       adsurl = {https://ui.adsabs.harvard.edu/abs/2019ApJ...886..137P}
}

@article{Chatterjee2020,
    author = {Chatterjee, Arka and Dutta, Broja G and Nandi, Prantik and Chakrabarti, Sandip K},
    title = "{Time-domain variability properties of XTE J1650−500 during its 2001 outburst: evidence of disc–jet connection}",
    journal = {\mnras},
    volume = {497},
    number = {4},
    pages = {4222-4230},
    year = {2020},
    month = {08},
     issn = {0035-8711},
    doi = {10.1093/mnras/staa2263},
    url = {https://doi.org/10.1093/mnras/staa2263},
    eprint = {https://academic.oup.com/mnras/article-pdf/497/4/4222/33671296/staa2263.pdf},
}

@ARTICLE{Blandford2019,
       author = {{Blandford}, Roger and {Meier}, David and {Readhead}, Anthony},
        title = "{Relativistic Jets from Active Galactic Nuclei}",
      journal = {\araa},
         year = 2019,
        month = aug,
       volume = {57},
        pages = {467-509},
          doi = {10.1146/annurev-astro-081817-051948},
archivePrefix = {arXiv},
       eprint = {1812.06025},
 primaryClass = {astro-ph.HE},
       adsurl = {https://ui.adsabs.harvard.edu/abs/2019ARA&A..57..467B}
}

@ARTICLE{Espi2020,
       author = {{Espinasse}, Mathilde and {Corbel}, St{\'e}phane and {Kaaret}, Philip and {Tremou}, Evangelia and {Migliori}, Giulia and {Plotkin}, Richard M. and {Bright}, Joe and {Tomsick}, John and {Tzioumis}, Anastasios and {Fender}, Rob and {Orosz}, Jerome A. and {Gallo}, Elena and {Homan}, Jeroen and {Jonker}, Peter G. and {Miller-Jones}, James C.~A. and {Russell}, David M. and {Motta}, Sara},
        title = "{Relativistic X-Ray Jets from the Black Hole X-Ray Binary MAXI J1820+070}",
      journal = {\apjl},
         year = 2020,
        month = jun,
       volume = {895},
       number = {2},
          eid = {L31},
        pages = {L31},
          doi = {10.3847/2041-8213/ab88b6},
archivePrefix = {arXiv},
       eprint = {2004.06416},
 primaryClass = {astro-ph.HE},
       adsurl = {https://ui.adsabs.harvard.edu/abs/2020ApJ...895L..31E}
}

@ARTICLE{MR1994,
       author = {{Mirabel}, I.~F. and {Rodr{\'\i}guez}, L.~F.},
        title = "{A superluminal source in the Galaxy}",
      journal = {\nat},
         year = 1994,
        month = sep,
       volume = {371},
       number = {6492},
        pages = {46-48},
          doi = {10.1038/371046a0},
       adsurl = {https://ui.adsabs.harvard.edu/abs/1994Natur.371...46M}
}

@ARTICLE{Berger2012,
       author = {{Berger}, E. and {Zauderer}, A. and {Pooley}, G.~G. and {Soderberg}, A.~M. and {Sari}, R. and {Brunthaler}, A. and {Bietenholz}, M.~F.},
        title = "{Radio Monitoring of the Tidal Disruption Event Swift J164449.3+573451. I. Jet Energetics and the Pristine Parsec-scale Environment of a Supermassive Black Hole}",
      journal = {\apj},
         year = 2012,
        month = mar,
       volume = {748},
       number = {1},
          eid = {36},
        pages = {36},
          doi = {10.1088/0004-637X/748/1/36},
archivePrefix = {arXiv},
       eprint = {1112.1697},
 primaryClass = {astro-ph.HE},
       adsurl = {https://ui.adsabs.harvard.edu/abs/2012ApJ...748...36B}
}

@ARTICLE{Tch2014,
       author = {{Tchekhovskoy}, Alexander and {Metzger}, Brian D. and {Giannios}, Dimitrios and {Kelley}, Luke Z.},
        title = "{Swift J1644+57 gone MAD: the case for dynamically important magnetic flux threading the black hole in a jetted tidal disruption event}",
      journal = {\mnras},
         year = 2014,
        month = jan,
       volume = {437},
       number = {3},
        pages = {2744-2760},
          doi = {10.1093/mnras/stt2085},
archivePrefix = {arXiv},
       eprint = {1301.1982},
 primaryClass = {astro-ph.HE},
       adsurl = {https://ui.adsabs.harvard.edu/abs/2014MNRAS.437.2744T}
}

@ARTICLE{Burrows2005,
       author = {{Burrows}, David N. and {Hill}, J.~E. and {Nousek}, J.~A. and {Kennea}, J.~A. and {Wells}, A. and {Osborne}, J.~P. and {Abbey}, A.~F. and {Beardmore}, A. and {Mukerjee}, K. and {Short}, A.~D.~T. and {Chincarini}, G. and {Campana}, S. and {Citterio}, O. and {Moretti}, A. and {Pagani}, C. and {Tagliaferri}, G. and {Giommi}, P. and {Capalbi}, M. and {Tamburelli}, F. and {Angelini}, L. and {Cusumano}, G. and {Br{\"a}uninger}, H.~W. and {Burkert}, W. and {Hartner}, G.~D.},
        title = "{The Swift X-Ray Telescope}",
      journal = {\ssr},
         year = 2005,
        month = oct,
       volume = {120},
       number = {3-4},
        pages = {165-195},
          doi = {10.1007/s11214-005-5097-2},
archivePrefix = {arXiv},
       eprint = {astro-ph/0508071},
 primaryClass = {astro-ph},
       adsurl = {https://ui.adsabs.harvard.edu/abs/2005SSRv..120..165B}
}

@ARTICLE{Bennett2003,
       author = {{Bennett}, C.~L. and {Halpern}, M. and {Hinshaw}, G. and {Jarosik}, N. and {Kogut}, A. and {Limon}, M. and {Meyer}, S.~S. and {Page}, L. and {Spergel}, D.~N. and {Tucker}, G.~S. and {Wollack}, E. and {Wright}, E.~L. and {Barnes}, C. and {Greason}, M.~R. and {Hill}, R.~S. and {Komatsu}, E. and {Nolta}, M.~R. and {Odegard}, N. and {Peiris}, H.~V. and {Verde}, L. and {Weiland}, J.~L.},
        title = "{First-Year Wilkinson Microwave Anisotropy Probe (WMAP) Observations: Preliminary Maps and Basic Results}",
      journal = {\apjs},
         year = 2003,
        month = sep,
       volume = {148},
       number = {1},
        pages = {1-27},
          doi = {10.1086/377253},
archivePrefix = {arXiv},
       eprint = {astro-ph/0302207},
 primaryClass = {astro-ph},
       adsurl = {https://ui.adsabs.harvard.edu/abs/2003ApJS..148....1B}
}

@ARTICLE{Singh1985,
       author = {{Singh}, K.~P. and {Garmire}, G.~P. and {Nousek}, J.},
        title = "{Observations of Soft X-Ray Spectra from a Seyfert 1 and a Narrow Emission-Line Galaxy}",
      journal = {\apj},
         year = 1985,
        month = oct,
       volume = {297},
        pages = {633},
          doi = {10.1086/163560},
       adsurl = {https://ui.adsabs.harvard.edu/abs/1985ApJ...297..633S}
}

@ARTICLE{Arnaud1985,
       author = {{Arnaud}, K.~A. and {Branduardi-Raymont}, G. and {Culhane}, J.~L. and
         {Fabian}, A.~C. and {Hazard}, C. and {McGlynn}, T.~A. and
         {Shafer}, R.~A. and {Tennant}, A.~F. and {Ward}, M.~J.},
        title = "{EXOSAT observations of a strong soft X-ray excess in MKN 841.}",
      journal = {\mnras},
         year = 1985,
        month = nov,
       volume = {217},
        pages = {105-113},
          doi = {10.1093/mnras/217.1.105},
       adsurl = {https://ui.adsabs.harvard.edu/abs/1985MNRAS.217..105A}
}

@INPROCEEDINGS{Gendreau2012,
       author = {{Gendreau}, Keith C. and {Arzoumanian}, Zaven and {Okajima}, Takashi},
        title = "{The Neutron star Interior Composition ExploreR (NICER): an Explorer mission of opportunity for soft x-ray timing spectroscopy}",
    booktitle = {Space Telescopes and Instrumentation 2012: Ultraviolet to Gamma Ray},
         year = 2012,
       editor = {{Takahashi}, Tadayuki and {Murray}, Stephen S. and {den Herder}, Jan-Willem A.},
       series = {Society of Photo-Optical Instrumentation Engineers (SPIE) Conference Series},
       volume = {8443},
        month = sep,
          eid = {844313},
        pages = {844313},
          doi = {10.1117/12.926396},
       adsurl = {https://ui.adsabs.harvard.edu/abs/2012SPIE.8443E..13G}
}

@ARTICLE{Tanaka1995,
       author = {{Tanaka}, Y. and {Nandra}, K. and {Fabian}, A.~C. and {Inoue}, H. and {Otani}, C. and {Dotani}, T. and {Hayashida}, K. and {Iwasawa}, K. and {Kii}, T. and {Kunieda}, H. and {Makino}, F. and {Matsuoka}, M.},
        title = "{Gravitationally redshifted emission implying an accretion disk and massive black hole in the active galaxy MCG-6-30-15}",
      journal = {\nat},
         year = 1995,
        month = jun,
       volume = {375},
       number = {6533},
        pages = {659-661},
          doi = {10.1038/375659a0},
       adsurl = {https://ui.adsabs.harvard.edu/abs/1995Natur.375..659T}
}

@ARTICLE{Martocchia1996,
       author = {{Martocchia}, Andrea and {Matt}, Giorgio},
        title = "{Iron Kalpha line intensity from accretion discs around rotating black holes}",
      journal = {\mnras},
         year = 1996,
        month = oct,
       volume = {282},
       number = {4},
        pages = {L53-L57},
          doi = {10.1093/mnras/282.4.L53},
       adsurl = {https://ui.adsabs.harvard.edu/abs/1996MNRAS.282L..53M}
}

@ARTICLE{HM1991,
       author = {{Haardt}, F. and {Maraschi}, L.},
        title = "{A Two-Phase Model for the X-Ray Emission from Seyfert Galaxies}",
      journal = {\apjl},
         year = 1991,
        month = oct,
       volume = {380},
        pages = {L51},
          doi = {10.1086/186171},
       adsurl = {https://ui.adsabs.harvard.edu/abs/1991ApJ...380L..51H}
}

@ARTICLE{Galeev1979,
       author = {{Galeev}, A.~A. and {Rosner}, R. and {Vaiana}, G.~S.},
        title = "{Structured coronae of accretion disks.}",
      journal = {\apj},
         year = 1979,
        month = apr,
       volume = {229},
        pages = {318-326},
          doi = {10.1086/156957},
       adsurl = {https://ui.adsabs.harvard.edu/abs/1979ApJ...229..318G}
}

@ARTICLE{VU2006,
       author = {{Vaughan}, Simon and {Uttley}, Philip},
        title = "{Detecting X-ray QPOs in active galaxies}",
      journal = {Advances in Space Research},
         year = 2006,
        month = jan,
       volume = {38},
       number = {7},
        pages = {1405-1408},
          doi = {10.1016/j.asr.2005.02.064},
archivePrefix = {arXiv},
       eprint = {astro-ph/0506456},
 primaryClass = {astro-ph},
       adsurl = {https://ui.adsabs.harvard.edu/abs/2006AdSpR..38.1405V}
}

@ARTICLE{Mummery2021,
       author = {{Mummery}, Andrew and {Balbus}, Steven A.},
        title = "{Hard X-ray emission from a Compton scattering corona in large black hole mass tidal disruption events}",
      journal = {\mnras},
         year = 2021,
        month = jul,
       volume = {504},
       number = {4},
        pages = {4730-4742},
          doi = {10.1093/mnras/stab1184},
archivePrefix = {arXiv},
       eprint = {2104.06195},
 primaryClass = {astro-ph.HE},
       adsurl = {https://ui.adsabs.harvard.edu/abs/2021MNRAS.504.4730M}
}

@ARTICLE{Alston2015,
       author = {{Alston}, W.~N. and {Parker}, M.~L. and {Markevi{\v{c}}i{\={u}}t{\.{e}}}, J. and {Fabian}, A.~C. and {Middleton}, M. and {Lohfink}, A. and {Kara}, E. and {Pinto}, C.},
        title = "{Discovery of an {\ensuremath{\sim}}2-h high-frequency X-ray QPO and iron K{\ensuremath{\alpha}} reverberation in the active galaxy MS 2254.9-3712}",
      journal = {\mnras},
         year = 2015,
        month = may,
       volume = {449},
       number = {1},
        pages = {467-476},
          doi = {10.1093/mnras/stv351},
archivePrefix = {arXiv},
       eprint = {1411.0684},
 primaryClass = {astro-ph.HE},
       adsurl = {https://ui.adsabs.harvard.edu/abs/2015MNRAS.449..467A}
}

@ARTICLE{Alston2020,
       author = {{Alston}, William N. and {Fabian}, Andrew C. and {Kara}, Erin and {Parker}, Michael L. and {Dovciak}, Michal and {Pinto}, Ciro and {Jiang}, Jiachen and {Middleton}, Matthew J. and {Miniutti}, Giovanni and {Walton}, Dominic J. and {Wilkins}, Dan R. and {Buisson}, Douglas J.~K. and {Caballero-Garcia}, Maria D. and {Cackett}, Edward M. and {De Marco}, Barbara and {Gallo}, Luigi C. and {Lohfink}, Anne M. and {Reynolds}, Chris S. and {Uttley}, Phil and {Young}, Andrew J. and {Zogbhi}, Abderahmen},
        title = "{A dynamic black hole corona in an active galaxy through X-ray reverberation mapping}",
      journal = {Nature Astronomy},
         year = 2020,
        month = jan,
       volume = {4},
        pages = {597-602},
          doi = {10.1038/s41550-019-1002-x},
archivePrefix = {arXiv},
       eprint = {2001.06454},
 primaryClass = {astro-ph.HE},
       adsurl = {https://ui.adsabs.harvard.edu/abs/2020NatAs...4..597A}
}

@INPROCEEDINGS{GR2014,
       author = {{Gonz{\'a}lez-Rodr{\'\i}guez}, A. and {Castro-Tirado}, A.~J. and {Guerrero}, M.~A. and {Castell{\'o}n}, A.},
        title = "{Late variability of flux and spectra of the tTidal disruption flare Sw J1644+57 from XMM-Newton data}",
    booktitle = {Revista Mexicana de Astronomia y Astrofisica Conference Series},
         year = 2014,
       series = {Revista Mexicana de Astronomia y Astrofisica Conference Series},
       volume = {45},
        month = dec,
        pages = {73},
archivePrefix = {arXiv},
       eprint = {1402.4057},
 primaryClass = {astro-ph.HE},
       adsurl = {https://ui.adsabs.harvard.edu/abs/2014RMxAC..45...73G}
}

@ARTICLE{Seifina2017,
       author = {{Seifina}, Elena and {Titarchuk}, Lev and {Virgilli}, Enrico},
        title = "{Swift J164449.3+573451 and Swift J2058.4+0516: Black hole mass estimates for tidal disruption event sources}",
      journal = {\aap},
         year = 2017,
        month = nov,
       volume = {607},
          eid = {A38},
        pages = {A38},
          doi = {10.1051/0004-6361/201730869},
archivePrefix = {arXiv},
       eprint = {1707.05898},
 primaryClass = {astro-ph.GA},
       adsurl = {https://ui.adsabs.harvard.edu/abs/2017A&A...607A..38S}
}

@ARTICLE{Eracleous1995,
       author = {{Eracleous}, Michael and {Livio}, Mario and {Halpern}, Jules P. and
         {Storchi-Bergmann}, Thaisa},
        title = "{Elliptical Accretion Disks in Active Galactic Nuclei}",
      journal = {\apj},
         year = 1995,
        month = jan,
       volume = {438},
        pages = {610},
          doi = {10.1086/175104},
       adsurl = {https://ui.adsabs.harvard.edu/abs/1995ApJ...438..610E}
}

@ARTICLE{Ricci2020,
       author = {{Ricci}, C. and {Kara}, E. and {Loewenstein}, M. and {Trakhtenbrot}, B. and
         {Arcavi}, I. and {Remillard}, R. and {Fabian}, A.~C. and
         {Gendreau}, K.~C. and {Arzoumanian}, Z. and {Li}, R. and {Ho}, L.~C. and
         {MacLeod}, C.~L. and {Cackett}, E. and {Altamirano}, D. and {Gand
        hi}, P. and {Kosec}, P. and {Pasham}, D. and {Steiner}, J. and
         {Chan}, C. -H.},
        title = "{The Destruction and Recreation of the X-Ray Corona in a Changing-look Active Galactic Nucleus}",
      journal = {\apjl},
         year = 2020,
        month = jul,
       volume = {898},
       number = {1},
          eid = {L1},
        pages = {L1},
          doi = {10.3847/2041-8213/ab91a1},
archivePrefix = {arXiv},
       eprint = {2007.07275},
 primaryClass = {astro-ph.HE},
       adsurl = {https://ui.adsabs.harvard.edu/abs/2020ApJ...898L...1R}
}

@ARTICLE{Ricci2021,
       author = {{Ricci}, C. and {Loewenstein}, M. and {Kara}, E. and {Remillard}, R. and {Trakhtenbrot}, B. and {Arcavi}, I. and {Gendreau}, K.~C. and {Arzoumanian}, Z. and {Fabian}, A.~C. and {Li}, R. and {Ho}, L.~C. and {MacLeod}, C.~L. and {Cackett}, E. and {Altamirano}, D. and {Gandhi}, P. and {Kosec}, P. and {Pasham}, D. and {Steiner}, J. and {Chan}, C. -H.},
        title = "{The 450 days X-ray monitoring of the changing-look AGN 1ES 1927+654}",
      journal = {arXiv e-prints},
         year = 2021,
        month = feb,
          eid = {arXiv:2102.05666},
        pages = {arXiv:2102.05666},
archivePrefix = {arXiv},
       eprint = {2102.05666},
 primaryClass = {astro-ph.HE},
       adsurl = {https://ui.adsabs.harvard.edu/abs/2021arXiv210205666R}
}

@ARTICLE{Merloni2015,
       author = {{Merloni}, A. and {Dwelly}, T. and {Salvato}, M. and {Georgakakis}, A. and
         {Greiner}, J. and {Krumpe}, M. and {Nandra}, K. and {Ponti}, G. and
         {Rau}, A.},
        title = "{A tidal disruption flare in a massive galaxy? Implications for the fuelling mechanisms of nuclear black holes}",
      journal = {\mnras},
         year = 2015,
        month = sep,
       volume = {452},
       number = {1},
        pages = {69-87},
          doi = {10.1093/mnras/stv1095},
archivePrefix = {arXiv},
       eprint = {1503.04870},
 primaryClass = {astro-ph.HE},
       adsurl = {https://ui.adsabs.harvard.edu/abs/2015MNRAS.452...69M}
}

@ARTICLE{2011Lodato,
       author = {{Lodato}, Giuseppe and {Rossi}, Elena M.},
        title = "{Multiband light curves of tidal disruption events}",
      journal = {\mnras},
         year = 2011,
        month = jan,
       volume = {410},
       number = {1},
        pages = {359-367},
          doi = {10.1111/j.1365-2966.2010.17448.x},
archivePrefix = {arXiv},
       eprint = {1008.4589},
 primaryClass = {astro-ph.CO},
       adsurl = {https://ui.adsabs.harvard.edu/abs/2011MNRAS.410..359L}
}

@ARTICLE{2012Merloni,
       author = {{Merloni}, A. and {Predehl}, P. and {Becker}, W. and {B{\"o}hringer}, H. and {Boller}, T. and {Brunner}, H. and {Brusa}, M. and {Dennerl}, K. and {Freyberg}, M. and {Friedrich}, P. and {Georgakakis}, A. and {Haberl}, F. and {Hasinger}, G. and {Meidinger}, N. and {Mohr}, J. and {Nandra}, K. and {Rau}, A. and {Reiprich}, T.~H. and {Robrade}, J. and {Salvato}, M. and {Santangelo}, A. and {Sasaki}, M. and {Schwope}, A. and {Wilms}, J. and {German eROSITA Consortium}, the},
        title = "{eROSITA Science Book: Mapping the Structure of the Energetic Universe}",
      journal = {arXiv e-prints},
         year = 2012,
        month = sep,
          eid = {arXiv:1209.3114},
        pages = {arXiv:1209.3114},
archivePrefix = {arXiv},
       eprint = {1209.3114},
 primaryClass = {astro-ph.HE},
       adsurl = {https://ui.adsabs.harvard.edu/abs/2012arXiv1209.3114M}
}

@ARTICLE{Lin2013,
       author = {{Lin}, Dacheng and {Irwin}, Jimmy A. and {Godet}, Olivier and {Webb}, Natalie A. and {Barret}, Didier},
        title = "{A \raisebox{-0.5ex}\textasciitilde 3.8 hr Periodicity from an Ultrasoft Active Galactic Nucleus Candidate}",
      journal = {\apjl},
         year = 2013,
        month = oct,
       volume = {776},
       number = {1},
          eid = {L10},
        pages = {L10},
          doi = {10.1088/2041-8205/776/1/L10},
archivePrefix = {arXiv},
       eprint = {1309.4440},
 primaryClass = {astro-ph.HE},
       adsurl = {https://ui.adsabs.harvard.edu/abs/2013ApJ...776L..10L}
}

@ARTICLE{Pasham2019,
       author = {{Pasham}, Dheeraj R. and {Remillard}, Ronald A. and {Fragile}, P. Chris and {Franchini}, Alessia and {Stone}, Nicholas C. and {Lodato}, Giuseppe and {Homan}, Jeroen and {Chakrabarty}, Deepto and {Baganoff}, Frederick K. and {Steiner}, James F. and {Coughlin}, Eric R. and {Pasham}, Nishanth R.},
        title = "{A loud quasi-periodic oscillation after a star is disrupted by a massive black hole}",
      journal = {Science},
         year = 2019,
        month = feb,
       volume = {363},
       number = {6426},
        pages = {531-534},
          doi = {10.1126/science.aar7480},
archivePrefix = {arXiv},
       eprint = {1810.10713},
 primaryClass = {astro-ph.HE},
       adsurl = {https://ui.adsabs.harvard.edu/abs/2019Sci...363..531P}
}

@ARTICLE{MR2009,
       author = {{McClintock}, Jeffrey E. and {Remillard}, Ronald A. and {Rupen}, Michael P. and {Torres}, M.~A.~P. and {Steeghs}, D. and {Levine}, Alan M. and {Orosz}, Jerome A.},
        title = "{The 2003 Outburst of the X-Ray Transient H1743-322: Comparisons with the Black Hole Microquasar XTE J1550-564}",
      journal = {\apj},
         year = 2009,
        month = jun,
       volume = {698},
       number = {2},
        pages = {1398-1421},
          doi = {10.1088/0004-637X/698/2/1398},
archivePrefix = {arXiv},
       eprint = {0705.1034},
 primaryClass = {astro-ph},
       adsurl = {https://ui.adsabs.harvard.edu/abs/2009ApJ...698.1398M}
}

@ARTICLE{Kara2019,
       author = {{Kara}, E. and {Steiner}, J.~F. and {Fabian}, A.~C. and {Cackett}, E.~M. and {Uttley}, P. and {Remillard}, R.~A. and {Gendreau}, K.~C. and {Arzoumanian}, Z. and {Altamirano}, D. and {Eikenberry}, S. and {Enoto}, T. and {Homan}, J. and {Neilsen}, J. and {Stevens}, A.~L.},
        title = "{The corona contracts in a black-hole transient}",
      journal = {\nat},
         year = 2019,
        month = jan,
       volume = {565},
       number = {7738},
        pages = {198-201},
          doi = {10.1038/s41586-018-0803-x},
archivePrefix = {arXiv},
       eprint = {1901.03877},
 primaryClass = {astro-ph.HE},
       adsurl = {https://ui.adsabs.harvard.edu/abs/2019Natur.565..198K}
}

@ARTICLE{Stein2021,
       author = {{Stein}, Robert and {Velzen}, Sjoert van and {Kowalski}, Marek and {Franckowiak}, Anna and {Gezari}, Suvi and {Miller-Jones}, James C.~A. and {Frederick}, Sara and {Sfaradi}, Itai and {Bietenholz}, Michael F. and {Horesh}, Assaf and {Fender}, Rob and {Garrappa}, Simone and {Ahumada}, Tom{\'a}s and {Andreoni}, Igor and {Belicki}, Justin and {Bellm}, Eric C. and {B{\"o}ttcher}, Markus and {Brinnel}, Valery and {Burruss}, Rick and {Cenko}, S. Bradley and {Coughlin}, Michael W. and {Cunningham}, Virginia and {Drake}, Andrew and {Farrar}, Glennys R. and {Feeney}, Michael and {Foley}, Ryan J. and {Gal-Yam}, Avishay and {Golkhou}, V. Zach and {Goobar}, Ariel and {Graham}, Matthew J. and {Hammerstein}, Erica and {Helou}, George and {Hung}, Tiara and {Kasliwal}, Mansi M. and {Kilpatrick}, Charles D. and {Kong}, Albert K.~H. and {Kupfer}, Thomas and {Laher}, Russ R. and {Mahabal}, Ashish A. and {Masci}, Frank J. and {Necker}, Jannis and {Nordin}, Jakob and {Perley}, Daniel A. and {Rigault}, Mickael and {Reusch}, Simeon and {Rodriguez}, Hector and {Rojas-Bravo}, C{\'e}sar and {Rusholme}, Ben and {Shupe}, David L. and {Singer}, Leo P. and {Sollerman}, Jesper and {Soumagnac}, Maayane T. and {Stern}, Daniel and {Taggart}, Kirsty and {van Santen}, Jakob and {Ward}, Charlotte and {Woudt}, Patrick and {Yao}, Yuhan},
        title = "{A tidal disruption event coincident with a high-energy neutrino}",
      journal = {Nature Astronomy},
         year = 2021,
        month = jan,
       volume = {5},
        pages = {510-518},
          doi = {10.1038/s41550-020-01295-8},
archivePrefix = {arXiv},
       eprint = {2005.05340},
 primaryClass = {astro-ph.HE},
       adsurl = {https://ui.adsabs.harvard.edu/abs/2021NatAs...5..510S}
}

@ARTICLE{Hayasaki2021,
       author = {{Hayasaki}, Kimitake},
        title = "{Neutrinos from tidal disruption events}",
      journal = {Nature Astronomy},
         year = 2021,
        month = jan,
       volume = {5},
        pages = {436-437},
          doi = {10.1038/s41550-021-01309-z},
archivePrefix = {arXiv},
       eprint = {2102.11879},
 primaryClass = {astro-ph.HE},
       adsurl = {https://ui.adsabs.harvard.edu/abs/2021NatAs...5..436H}
}

@ARTICLE{NW2013,
       author = {{Niko{\l}ajuk}, M. and {Walter}, R.},
        title = "{Tidal disruption of a super-Jupiter by a massive black hole}",
      journal = {\aap},
         year = 2013,
        month = apr,
       volume = {552},
          eid = {A75},
        pages = {A75},
          doi = {10.1051/0004-6361/201220664},
archivePrefix = {arXiv},
       eprint = {1304.0397},
 primaryClass = {astro-ph.HE},
       adsurl = {https://ui.adsabs.harvard.edu/abs/2013A&A...552A..75N}
}

@ARTICLE{VU2005,
       author = {{Vaughan}, S. and {Uttley}, P.},
        title = "{Where are the X-ray quasi-periodic oscillations in active galaxies?}",
      journal = {\mnras},
         year = 2005,
        month = sep,
       volume = {362},
       number = {1},
        pages = {235-244},
          doi = {10.1111/j.1365-2966.2005.09296.x},
archivePrefix = {arXiv},
       eprint = {astro-ph/0506455},
 primaryClass = {astro-ph},
       adsurl = {https://ui.adsabs.harvard.edu/abs/2005MNRAS.362..235V}
}

@ARTICLE{Gierlinski2008,
       author = {{Gierli{\'n}ski}, Marek and {Middleton}, Matthew and {Ward}, Martin and {Done}, Chris},
        title = "{A periodicity of \raisebox{-0.5ex}\textasciitilde1hour in X-ray emission from the active galaxy RE J1034+396}",
      journal = {\nat},
         year = 2008,
        month = sep,
       volume = {455},
       number = {7211},
        pages = {369-371},
          doi = {10.1038/nature07277},
       adsurl = {https://ui.adsabs.harvard.edu/abs/2008Natur.455..369G}
}

@ARTICLE{Ashton2021,
       author = {{Ashton}, Dominic I. and {Middleton}, Matthew J.},
        title = "{Searching for energy-resolved quasi-periodic oscillations in AGN}",
      journal = {\mnras},
         year = 2021,
        month = mar,
       volume = {501},
       number = {4},
        pages = {5478-5499},
          doi = {10.1093/mnras/staa4024},
archivePrefix = {arXiv},
       eprint = {2101.01194},
 primaryClass = {astro-ph.HE},
       adsurl = {https://ui.adsabs.harvard.edu/abs/2021MNRAS.501.5478A}
}

@ARTICLE{Magano2016,
       author = {{Mangano}, V. and {Burrows}, D.~N. and {Sbarufatti}, B. and {Cannizzo}, J.~K.},
        title = "{The Definitive X-Ray Light Curve of Swift J164449.3+573451}",
      journal = {\apj},
         year = 2016,
        month = feb,
       volume = {817},
       number = {2},
          eid = {103},
        pages = {103},
          doi = {10.3847/0004-637X/817/2/103},
archivePrefix = {arXiv},
       eprint = {1511.06447},
 primaryClass = {astro-ph.HE},
       adsurl = {https://ui.adsabs.harvard.edu/abs/2016ApJ...817..103M}
}

@ARTICLE{Zauderer2011,
       author = {{Zauderer}, B.~A. and {Berger}, E. and {Soderberg}, A.~M. and {Loeb}, A. and {Narayan}, R. and {Frail}, D.~A. and {Petitpas}, G.~R. and {Brunthaler}, A. and {Chornock}, R. and {Carpenter}, J.~M. and {Pooley}, G.~G. and {Mooley}, K. and {Kulkarni}, S.~R. and {Margutti}, R. and {Fox}, D.~B. and {Nakar}, E. and {Patel}, N.~A. and {Volgenau}, N.~H. and {Culverhouse}, T.~L. and {Bietenholz}, M.~F. and {Rupen}, M.~P. and {Max-Moerbeck}, W. and {Readhead}, A.~C.~S. and {Richards}, J. and {Shepherd}, M. and {Storm}, S. and {Hull}, C.~L.~H.},
        title = "{Birth of a relativistic outflow in the unusual {\ensuremath{\gamma}}-ray transient Swift J164449.3+573451}",
      journal = {\nat},
         year = 2011,
        month = aug,
       volume = {476},
       number = {7361},
        pages = {425-428},
          doi = {10.1038/nature10366},
archivePrefix = {arXiv},
       eprint = {1106.3568},
 primaryClass = {astro-ph.HE},
       adsurl = {https://ui.adsabs.harvard.edu/abs/2011Natur.476..425Z}
}

@ARTICLE{Bloom2011a,
       author = {{Bloom}, J.~S. and {Butler}, N.~R. and {Cenko}, S.~B. and {Perley}, D.~A.},
        title = "{GRB 110328A / Swift J164449.3+573451: X-ray analysis and a mini-blazar  analogy.}",
      journal = {GRB Coordinates Network},
         year = 2011,
        month = mar,
       volume = {11847},
        pages = {1},
       adsurl = {https://ui.adsabs.harvard.edu/abs/2011GCN.11847....1B}
}
\bibliographystyle{aasjournal}

\end{document}